\documentclass[12pt, prd]{revtex4}
\usepackage{graphicx,epsfig}
\input{epsf.tex}
\usepackage{amsmath}
\usepackage{amssymb}

\usepackage[utf8]{inputenc}

\begin{document}

\title{When spacetime vibrates: An introduction to gravitational waves}
\author{Jos\'{e} P. S. Lemos}
\affiliation{Centro de Astrof\'isica e Gravita\c{c}\~ao - CENTRA,
Departamento de F\'isica, Instituto Superior T\'{e}cnico - IST,
Universidade de Lisboa - UL, Avenida Rovisco Pais 1, 1049-001, Lisboa,
Portugal, Electronic address: joselemos@tecnico.ulisboa.pt}



\begin{abstract}
This article presents a comprehensive analysis of the physics of
gravitational waves, exploring both the theoretical foundations and
the most recent experimental advances. After a general introduction to
the theory of general relativity and its major implications, the
article discusses the history of gravitational waves,
from their prediction by Einstein to their actual detection. It then
explains what gravitational waves are and how they interact with
appropriate detectors.
The main mechanisms of gravitational radiation emission are analyzed,
with a focus on compact binary systems of compact objects, whose orbits
typically evolve in three phases: inspiral, merger, and the final
ringdown phase, each of these phases leaving distinct signatures in
the emitted waves.
The article highlights the fundamental role of the giant
interferometers LIGO, Virgo, and KAGRA, true cathedrals of modern
science, and revisits the historic event GW150914, the first direct
detection of gravitational waves, which confirmed the predictions of
general relativity and opened a new era for astronomy. This
achievement was recognized with the 2017 Nobel Prize in Physics.
Other observed events are also discussed, along with their
astrophysical sources, and the possibility of detecting gravitational
waves of cosmological origin, originating from the Big Bang
itself. Finally, current and future projects are analyzed, including
observatories based on increasingly sophisticated interferometers, as
well as proposals for alternative detection methods, illustrating how
gravitational-wave astronomy is shaping the present and future of our
exploration of the universe.
In concluding, the detection of gravitational waves is set in a
broader context by examining the discoveries across the
electromagnetic spectrum, thereby illustrating the complementary
perspectives these different observational channels provide.

\end{abstract}
\maketitle

\section{General Relativity and its major consequences}

General relativity changed the world and our view of the world.
The fundamental equation is
\begin{equation}
G_{ab}=\frac{8\pi G}{c^4}T_{ab} .
\label{Eequation}
\end{equation}
Here,
$G_{ab}$ represents spacetime and $T_{ab}$ represents matter, with the
two quantities coupled through the universal gravitational constant
$G$ and the speed of light $c$ in the precise combination
$\frac{8\pi
G}{c^4}$. The quantity $G_{ab}$, known as the Einstein tensor, is
constructed from the spacetime metric and its
first and second derivatives with respect
to space and time.
Equation~\eqref{Eequation}
expresses the profound idea
that spacetime tells matter how to move, and
matter tells spacetime how to curve. It replaced the Newtonian concept
of gravity, namely, a universal gravitational force between masses,
with a geometric description of gravity based on the concept of
spacetime.
Proposed by Einstein in 1915,
Eq.~\eqref{Eequation}, called the Einstein equation,
forms the foundation of general relativity.

The theory opened up entirely new and important branches of science:

-    Gravitational lensing (1915, Einstein)

-    Gravitational waves (1916, Einstein)

-    Cosmology (1917, Einstein)

-    Fundamental theories (1918, Weyl)

-    Black holes (1939, Oppenheimer and Snyder)

Gravitational lenses are massive objects that bend and focus light
from distant sources through their gravitational fields, producing
brighter, distorted, or multiple images compared to direct
observation.
The phenomenon was initially discussed by Einstein as
early as 1913 and formulated more rigorously within the
framework of general relativity in 1915.
The related effect of light deflection was famously confirmed
during the 1919 solar eclipse. 
Today, the study
of gravitational lensing is highly relevant in astrophysics and
cosmology and is a powerful tool for investigating the distribution
of matter in the universe, including dark matter.

Gravitational waves, ripples in spacetime that propagate at the speed
of light, were proposed by Einstein in 1916 shortly after the
formulation of general relativity. We will discuss these in more
detail.

Cosmology, as a branch of modern science, was inaugurated by Einstein
in 1917 with a remarkable paper in which he proposed a finite and closed
universe without boundary, revolutionizing ideas about physical
topology and the very concept of universe.

Fundamental theories aim to describe all of physics through a single
unifying law that integrates every physical  interaction. One of the
earliest attempts in this direction was made in 1918 by Weyl,
a German
mathematician and physicist, who proposed a unified
theory of gravitation and electromagnetism.

Black holes are objects formed purely of gravitational
fields. Their existence was first speculated in 1939 in an
extraordinary work by Oppenheimer and Snyder. Today, we know that
black holes are ubiquitous throughout the universe.

\section{History of gravitational waves}

\subsection{Pre-history}

The pre-history of gravitational waves is brief but interesting. Below
we highlight a few dates, names, and ideas:

\vskip 0.1cm
\noindent
1825 - Laplace, a French mathematician and physicist, speculated
    that the speed of gravity must be at least 100 million times
    faster than the speed of light.

\vskip 0.1cm
\noindent
1893 - Heaviside, an English physicist, proposed a formulation of
    gravity analogous to electromagnetism, anticipating concepts that
    would later appear in the theory of relativity, including the idea
    of gravitational waves.

\vskip 0.1cm
\noindent
1905 - Poincaré, a French mathematician and physicist,
while developing a
    relativistic theory of gravity inspired by Maxwell's equations,
    explicitly introduced the term gravitational wave.

\subsection{History}

The history of gravitational waves spans over a century. We briefly
characterize the key developments by listing the dates, the physicists
involved, and the contributions:

\vskip 0.1cm
\noindent
1916 - Einstein derived a formula for gravitational radiation. It
contained a major error. The radiation was monopolar, i.e.,
originating from an object expanding and contracting radially. Abraham,
a German physicist, and Nordstr\"om, a Finnish physicist, pointed
Einstein
in the correct direction, but they themselves did not derive the right
formula.

\vskip 0.1cm
\noindent
1918 - Einstein derived the correct quadrupole radiation formula for
gravitational wave emission, or almost correct, as it was missing a
factor of 2.

\vskip 0.1cm
\noindent
1922 - Eddington, an English astrophysicist, identified the error and
finally presented the correct formula with the missing factor of
2. He also argued that if longitudinal gravitational waves were real,
they would propagate at the speed of thought, i.e., instantaneously,
which would contradict the principles of relativity. From this, he
concluded that if gravitational waves do exist, only transverse ones
could be physically possible. He then applied the quadrupole formula
to a rotating bar, establishing a theoretical foundation for
calculating energy loss via gravitational radiation in astrophysical
systems.

\vskip 0.1cm
\noindent
1936 - Now at Princeton, Einstein and his collaborator Rosen, an
American
physicist, submitted a paper to Physical Review asserting that
gravitational waves could not occur. The manuscript was rejected after
an anonymous referee pointed out errors and suggested
corrections. Outraged that the paper had been reviewed without his
consent, Einstein wrote a protest letter to the editor, Tate,
vowing never to publish in that journal again. Shortly afterward, the
referee, Robertson, an American physicist at Caltech, discussed
the errors with Einstein's collaborator Infeld,
from Poland. Convinced by
Infeld, Einstein redid the calculations, acknowledged the mistake, and
a corrected version of the paper was published in the Journal of the
Franklin Institute.

\vskip 0.1cm
\noindent
1941 - Landau and Lifshitz, Russian physicists, showed that the
quadrupole formula is valid for describing the emission of
gravitational waves by binary star systems in orbit. This marked a
significant advance, building on Eddington's earlier work on rotating
bars.

\vskip 0.1cm
\noindent
1957 - Pirani, an English physicist, recognized that gravitational waves
are manifestations of spacetime curvature, described by the Riemann
tensor. He demonstrated that free-falling particles experience
deviations in their geodesic paths as a gravitational wave passes,
revealing the physically observable tidal-force nature of the
phenomenon. Newton was the first to understand tidal forces as
differential gravitational forces responsible for ocean tides.
Later, Einstein clarified that, within the framework of general
relativity, the gravitational force itself has no fundamental physical
meaning. Instead, tidal forces, manifested as spacetime
curvature and mathematically described by the Riemann tensor,
are the quantities that carry true physical
significance.
Building 
on this understanding, Bondi, an Austrian-born British physicist, showed
that gravitational waves carry energy as they propagate, a conclusion
reinforced by Feynman, an American physicist, who had
previously successfully quantized the electromagnetic field.

\vskip 0.1cm
\noindent
1962 - Bondi formally proved that systems emitting gravitational waves
lose mass, following work initiated in 1958 by the Polish
physicist Trautman.

\vskip 0.1cm
\noindent
1967 - Thorne, who had previously graduated from Princeton under the
supervision of Wheeler, both American physicists, began
systematically developing the theory of gravitational waves, laying
the foundation for its application to realistic astrophysical
contexts. In parallel, Penrose and Newman, English and American
physicists, respectively, introduced new mathematical tools, including
the Newman-Penrose formalism, which proved important for accurately
describing gravitational radiation in asymptotically flat spacetimes.

\vskip 0.1cm
\noindent
1969 - Burke, then a student of Thorne at Caltech, applied the method
of matched asymptotic expansions to general relativity, enabling
consistent matching between the near field of the gravitational-wave
source and the far-field region where radiation is observed.

\vskip 0.1cm
\noindent
1970 - Weber, an American experimental physicist,
used resonant bar detectors and
announced the detection of gravitational waves. However, his results
were never independently reproduced, and the scientific community
ultimately rejected the claim.

\vskip 0.1cm
\noindent
1971 - Hawking, an English physicist, used his area theorem, which
states that the event horizon area of black holes cannot decrease, to
show that in a collision of two black holes, there is a maximum amount
of mass that can be converted into gravitational radiation. For the
merger of two equal-mass, non-spinning black holes, this theoretical
limit is about 29\% of the total initial mass.

\vskip 0.1cm
\noindent
1974 - Hulse and Taylor, American astronomers at Princeton, discovered
a pulsar in a compact binary system and, over the following years,
showed that the system's orbital radius was shrinking in agreement
with energy loss via gravitational waves, as predicted by general
relativity. For this indirect confirmation of the existence of
gravitational waves, they received the 1993 Nobel Prize in Physics.

\vskip 0.1cm
\noindent
1975 - Weiss, a German-born American physicist of MIT, and Thorne met
in Washington, D.C., to discuss and develop the proposal for a
laser-based Michelson interferometer for detecting gravitational
waves, an idea originally conceived by Weiss in the 1960s. The
U.S. National Science Foundation, whose principal reviewer Itzikson
was an expert on gravitational waves, approved initial funding for the
project that would become LIGO, the Laser Interferometer
Gravitational-Wave Observatory. Soon after, Drever, a Scottish
experimental physicist, joined them, forming the troika with Weiss and
Thorne.

\vskip 0.1cm
\noindent
1991 - Christodoulou, a Greek physicist, mathematically demonstrated the
existence of the gravitational wave memory effect: a
permanent change in the relative separation of test particles after a
wave passes. Unlike the transient oscillation of the wave itself, this
effect of much smaller amplitude leaves a lasting imprint on spacetime.

\vskip 0.1cm
\noindent
1994 - Barry Barish, an American physicist, assumed leadership of the
LIGO project, introducing a renewed emphasis on experimental
feasibility and the practical construction of the detectors. Under his
direction, construction began on two LIGO observatories: one in
Livingston, Louisiana, and the other in Hanford, Washington.

\vskip 0.1cm
\noindent
2002 - LIGO began its first observational run, named O1, aiming to
detect gravitational waves. The Virgo interferometer, located near
Pisa, Italy, was later added to the Hanford and Livingston LIGO
interferometers, marking the beginning of international collaboration
in the search for gravitational waves.

\begin{figure*}[h]
\centering
\includegraphics[scale=0.65]{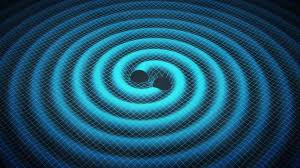}
\caption{
Representation of two black holes inspiraling on a collision
course, emitting gravitational waves, with the end point
being their merger.}
\label{twobhsandwaves}
\end{figure*}

\vskip 0.1cm
\noindent
2005 -
Pretorius, a South African physicist at Princeton, achieved the first
successful numerical relativity solution of the black hole merger
problem, see Fig.~\ref{twobhsandwaves}. His breakthrough inaugurated
a new era in the field, catalyzing the development of
gravitational wave templates now in use in modern interferometers.

\vskip 0.1cm
\noindent
2015 - On September 14, gravitational waves were detected for the
first time by LIGO, marking a historic milestone in physics. The
signal, named GW150914, was generated by the merger of two black
holes, each with a mass of about 30 solar masses, over 300 Mpc
away. Since 1 Mpc equals one million parsecs, and one parsec is about
3.3 light-years, the event occurred over a billion light-years from
Earth, and a billion years ago. The discovery confirmed Einstein's
general relativity prediction and initiated the era of gravitational
wave astronomy.

\vskip 0.1cm
\noindent
2017 - In October, in recognition of the historic detection of
gravitational waves, the Nobel Prize in Physics was awarded to Weiss,
Barish, and Thorne, three key figures in the development and
leadership of the LIGO project.

\vskip 0.1cm
\noindent
2020 - The Nobel Prize in Physics was awarded for fundamental
contributions to the understanding of black holes, whose existence was
confirmed directly via gravitational wave detections and, earlier,
indirectly via astrophysical observations. Half of the prize went to
Penrose, an English physicist and mathematician, for
proving, based on general relativity, that black hole
formation is an inevitable result of gravitational collapse. The other
half was shared by Genzel and Ghez, Germans and American
astronomers, respectively,
for their observational studies of the supermassive black
hole at the center of our galaxy. Although not directly tied to
gravitational wave detection, the prize highlighted the role of black
holes as key sources of these signals. Since the first
detection in 2015, black hole mergers have become central to
gravitational wave astronomy, reinforcing the significance of the
awarded work.

\vskip 0.1cm
\noindent
2023 - The first evidence of a low-frequency gravitational wave
background was announced, detected through the analysis of signals
from millisecond pulsars. This background is believed to originate
from
mergers of supermassive black holes throughout cosmic history.

\section{What is a gravitational wave?}

\subsection{Gravitational waves and their effect on detectors}

A gravitational wave is a vibration of spacetime that propagates at
the speed of light. When this vibration is sufficiently small, it can
be treated as a weak perturbation over a flat background, 
called Minkowski spacetime. A spacetime is characterized by a
metric denoted by $g_{ab}(x)$, where  $x$ generically represents the
four coordinates  $x^a$ of spacetime. Here, $x^0$
is reserved for the
time coordinate, and $x^1$, $x^2$, and $x^3$ for the spatial
coordinates. The indices $a,b$
appearing in $g_{ab}(x)$ denote the
various components of the metric, with $a,b=0,1,2,3$.  Let us
begin, then, with the metric for a gravitational wave,
\begin{equation}
g_{ab}(x)=\eta_{ab}+h_{ab}(x),
\label{gwmetric}
\end{equation}
where $\eta_{ab}={\rm diag}[-1,1,1,1]$
is the flat Minkowski metric with the off-diagonal
terms being zero, and $h_{ab}(x)$ is a small metric perturbation that
represents the gravitational wave, with components much smaller than
unity. Let us assume that the wave propagates in time along the
$z$-direction. By making appropriate choices for $h_{ab}(x)$, one can
show that Einstein's vacuum equations, $G_{ab} = 0$, see
Eq.~\eqref{Eequation}, lead, for the metric in the form of
Eq.~\eqref{gwmetric}, to the equation
\begin{equation}
\Box h_{ab}=0,
\label{dalmbwertianhab}
\end{equation}
where $\Box h_{ab}$ denotes the d'Alembertian of $h_{ab}$, i.e.,
$\Box
h_{ab}=-\frac{1}{c^2}\frac{\partial^2 h_{ab}}{\partial t^2}
+\frac{\partial^2 h_{ab}}{\partial z^2}$,
and we are taking $x^0 = ct$, $x^3 = z$, with $t$ being time.
We will also write $x^1 = x$, $x^2 = y$. This is a wave equation
for a perturbation that propagates at speed $c$. Any solution of the
wave equation generically has the form $h\left(t -
\frac{z}{c}\right)$. Let us choose a simple form for $h$, namely,
$ h\left(t-\frac{z}{c}\right)=h\,{\sin}\,\omega\left(t-\frac{z}{c}\right)$.
It can be shown in this case, using Einstein's equations and the
properties of a wave, that the sixteen components of $h_{ab}$ are
reduced to just two independent components. One solution provides
$h_{xx}$ and $h_{yy}$ as the independent components. The gravitational
wave is therefore transverse. Moreover, the solution has the form
$h_{xx} = -h_{yy}$. We set
$h_{xx}=-h_{yy}=h\,{\sin}\,\omega\left(t-\frac{z}{c}\right)$.
In compact form, the solution for $h_{ab}$ is represented by a
$4\times 4$ matrix as follows,
\begin{equation}
h_{ab}(x)=\begin{pmatrix}
0&0&0&0\\
0&h&0&0\\
0&0&-h&0\\
0&0&0&0
\end{pmatrix}
{\sin}\,\omega\left(t-\frac{z}{c}\right).
\label{hab+matrix}
\end{equation}
The line element $ds^2$, which gives the square of infinitesimal
intervals and distances, defined as $ds^2 = g_{ab} dx^a dx^b$ with
$g_{ab}$ given in Eq.~\eqref{gwmetric} and $h_{ab}$ in
Eq.~\eqref{hab+matrix}, becomes in this case,
\begin{equation}
ds^2= -dt^2+\left[1+h\,{\sin}\,
\omega\left(t-\frac{z}{c}\right)
\right]dx^2
+\left[1-
h\,{\sin}\,      \omega\left(t-\frac{z}{c}\right)
\right]dy^2+dz^2.
\label{ds2dehab+matrix}
\end{equation}
The metric and the corresponding line element represent the spacetime
of a gravitational wave with amplitude $h$ and frequency $\omega$
propagating in the $z$-direction at the speed of light $c$.

\begin{figure*}[h]
\centering
\includegraphics[scale=1.35]{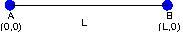}
\caption{
Two test particles, $A$ and
$B$, are separated by a distance $L$ along the $x$-axis.
As a gravitational wave passes, the particles will oscillate
according to the wave.}
\label{testparticlesgw}
\end{figure*}

The wave propagates in the $z$-direction, and there are two test
particles at points $A$ and  $B$, at rest waiting for the wave. Particle
$A$  has coordinates  $(x=0,y=0)$, and particle $B$ has coordinates
$(x=L,y=0)$, see Fig.~\ref{testparticlesgw}.
We assume that the  $z$-coordinate of the particles is given by $z=0$.

\begin{figure*}[h]
\centering
\includegraphics[scale=0.9]{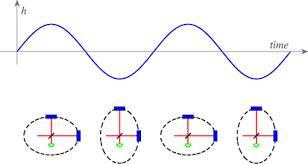}
\caption{
The upper part of the figure shows the deformation $h$ as a function of
time when the gravitational wave passes by the test particles placed
along the $x$-axis or the $y$-axis.  The lower part shows that the
particles oscillate in the $x\times y$ plane
with $+$ polarization.
}
\label{+deformation+}
\end{figure*}

When, by assumption, the wave passes perpendicularly to the line
joining the two particles, the proper distance between them varies
with time as $L'(t)=\int_0^L \left[1+h_{xx}(t,0)\right]^{\frac12}\,
dx$.
The waves that pass through a detector are very weak, so that
$h_{xx}\ll1$, and we can expand the square root and perform the
integral. Thus, the proper distance between the two particles
satisfies
$L'(t)=L \left[1+\frac12 h_{xx}(t,0)\right]$.
Therefore, the distance between $A$ and $B$ undergoes a
relative change in length given by
$\frac{\Delta L}{L}= \frac{L'-L}{L}=\frac12 h_{xx}(t,0)$.
Since
$h_{xx}(t,0)=h\,{\sin}\, \omega t$, we have
\newline
\begin{equation}
\frac{\Delta L}{L}=\frac12\,h\,{\sin}\, \omega t\,.
\label{DeltaL}
\end{equation}
\centerline{}
We see that there is a differential variation in length between
neighboring points along the $x$-direction. A differential variation in
length between neighboring points means a tidal change, and thus
gravitational waves carry tidal forces.
Along the $y$-direction, the proper distance between two
particles varies with time as
$L'(t)=L \left[1+\frac12 h_{yy}(t,0)\right]$.
Since $h_{yy}=-h_{xx}$, along $y$ we have
$\frac{\Delta L}{L}=-\frac12\,h\,{\sin}\, \omega t$,
assuming here that two other particles are placed along the $y$-axis.
Thus, test particles move in the plane perpendicular to the direction
of propagation of the wave, $z$, expanding in one direction while
contracting in the other, and then reversing again,
see
Fig.~\ref{+deformation+}.
This is called the $+$ solution,
the wave has $+$ polarization.
Note that the deformation is given by the maximum value of
$\frac{\Delta L}{L}$, i.e.,
$\frac12 h=\left(\frac{\Delta L}{L}\right)_{\rm max}$.
Additionally, note that for a given $h$, a property of the wave, the
larger the value of $L$,  a given parameter of the experiment, the
larger the corresponding $\Delta L$,
and so, measuring the variation $L$
itself becomes easier.

\newpage

There is yet another solution, independent of the $+$ solution, which
is given by $h_{xy}=h_{yx}=h\,{\sin}\,\omega\left(t-\frac{z}{c}\right)$,
and the
other components are zero. In this case, we write this solution for
$h_{ab}$
in the form
\begin{equation}               
h_{ab}(x)=\begin{pmatrix}
0&0&0&0\\
0&0&h&0\\
0&h&0&0\\
0&0&0&0
\end{pmatrix}
{\sin}\,      \omega\left(t-\frac{z}{c}\right).
\label{habx}
\end{equation}
It is called the $\times$ solution because of the 45$^{\rm o}$ rotation of the
polarization axes. It is the wave with $\times$ polarization. There are no
other independent solutions.

We can combine the two solutions, remembering that they are
independent and therefore the $h$ of one is different from the $h$ of
the other. For the $+$ solution we denote it by $h_+$, and for the $\times$
solution we denote it by
$h_\times$.
The general solution is then
\begin{equation}               
h_{ab}(x)=\begin{pmatrix}
0&0&0&0\\
0&h_+&h_\times&0\\
0&h_\times&-h_+&0\\
0&0&0&0
\end{pmatrix}
 {\sin}\,   \omega\left(t-\frac{z}{c}\right).
\label{hab+x}
\end{equation}
The gravitational wave has the property of being transverse, since it
propagates in the $z$ direction and has components only in the $x$
and $y$ directions, and of being traceless, since the trace is
zero,
$ h_{tt}+h_{xx}+h_{yy}+h_{zz}=0+h_++(-h_+)+0=0$.

We still need to connect the quantities detected in the detector with
the source emitting the gravitational waves.

\subsection{Gravitational radiation from a source}

\begin{figure*}[h]
\centering
\includegraphics[scale=0.5]{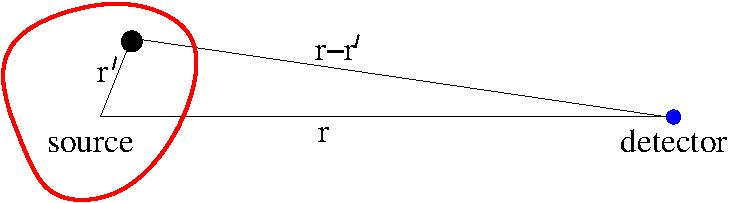}
\caption{
Representation of a source of gravitational waves and the
corresponding detector. A small element of the source is located at
$r'$, where $r'$ represents the position vector of that element
relative to a given origin. The position vector of the detector
relative to the origin is $r$. The position vector from the element of
the source to the detector is $r-r'$. The distances are obtained by
taking the magnitude of the position vectors. Each element of the
source is characterized by the quantity $T_{ab}$, which gives the energy
and momenta of that element.}
\label{farfieldenglishfigure}
\end{figure*}

Let us suppose a source of gravitational waves and a detector. The
origin of coordinates is located at some point inside the source. A
generic element of the source has position vector $r^\prime$ relative to the
origin, and the detector has position vector $r$, also relative to the
origin, see the diagram in Fig.~\ref{farfieldenglishfigure}.
Einstein's equation, see Eq.~\eqref{Eequation}, for weak fields with
a source, i.e., with some form of matter or energy, gives
$\Box\, h_{ab}=\frac{16 \pi G}{c^4}\,\left(T_{ab}-\frac12
g_{ab}{T^{c}}_c\right)$, 
where the metric form from Eq.~\eqref{gwmetric} has been used, and
where ${T^{c}}_c$ is the trace of $T_{ab}$.
Taking the trace of the preceding equation gives ${T^{c}}_c=0$ for this
case, and the equation then takes on the simpler form
\begin{equation}  
\Box\, h_{ab}=\frac{16 \pi G}{c^4}\,T_{ab}.
\label{habeqcao2}
\end{equation}

The formal solution of Eq.~\eqref{habeqcao2}
is known and given by an integral over
the source, with respect to $r^\prime$, i.e.,
$h_{ab}(r,t)=\frac{1}{4\pi}\frac{16 \pi G}{c^4} \int
\frac{1}{|r-r'|}T_{ab}\left(r',t-\frac{|r-r'|}{c}\right)\,d^3 x'$,
where
$h_{ab}(r,t)$
is the perturbation measured at the detector's position $r$ 
and at the detector's time $t$.
The time $t-\frac{|r-r'|}{c}$ is the retarded time, relating
events that occurred at the source with the detector's own time $t$.
For fields far away,  $r\gg r'$, one obtains
\begin{equation}  
h_{ab}(r,t)=\frac{4G}{c^4r}
 \int T_{ab}\left(r',t-\frac{r}{c}\right)\,d^3 x'.
\label{habintegrado}
\end{equation}
$T^{ab}$
is the important quantity for establishing the connection between the
source and the detector. It is the energy-momentum tensor because it
contains information about the energy density and the momentum density
of matter. In closed systems, energy and momentum obey conservation
laws, and thus $T^{ab}$ itself satisfies a conservation law.
For weak gravitational fields in the source, as we are assuming here,
the conservation law is ${T^{ab}}_{,b}=0$,
where, for simplicity, a comma denotes an ordinary derivative, i.e.
${T^{ab}}_{,b}\equiv\frac{\partial
T^{ab}}{\partial x^b}$
and, moreover, whenever indices are repeated, one sums over them,
this is the
Einstein summation convention. Since $a,b=0,1,2,3$,
we have, by this convention,
${T^{ab}}_{,b}={T^{a0}}_{,0}+{T^{ak}}_{,k}$,
with $k=1,2,3$ now being the summed spatial index. From the
conservation law we then have ${T^{a0}}_{,0}=-{T^{ak}}_{,k}$.
Setting the index $a=0$, gives
${T^{00}}_{,0}=-{T^{0k}}_{,k}$.
Integrating, we obtain
$\frac{1}{c}\frac{\partial }{\partial t} \int T^{00} d^3x=-\int
\frac{\partial T^{0k} }{\partial x^k}d^3x$,
where the integral is taken over space.
Using the divergence theorem, the last term becomes
$\int
\frac{\partial T^{0k} }{\partial x^k}d^3x=\int T^{0k} d\Sigma_k$
where $\Sigma_k$
is a surface that encloses the source. But on this surface
there is no longer any source, so $T^{0k}=0$. Therefore,
 $\frac{1}{c}\frac{\partial
}{\partial t} \int T^{00} d^3x=0$
which implies that
 $\int T^{00}
d^3x={\rm constant}$.
Similarly, one finds that $\int T^{0k}
d^3x={\rm constant}$.
Thus, the terms  $T^{00}$ and $T^{0k}$
are time independent,
and since we want
$h_{ab}$ to be a
time-dependent wave-like perturbation, we put in
advance $h_{00}=0$ and $h_{0k}=0$.

The remaining terms $T^{ij}$, with $i,j=1,2,3$, give a contribution to
$h_{ab}$, i.e., to $h_{ij}$. Let us see this. We use again the
expression of the conservation law, i.e.,
${T^{a0}}_{,0}=-{T^{aj}}_{,j}$.
Setting the index $a=0$ and taking the time derivative of this
expression, we obtain ${T^{00}}_{,00}=-{T^{0j}}_{,0j}$, where we have
used that partial derivatives commute. On the other hand, for $a=i$,
we have ${T^{i0}}_{,0}=-{T^{ij}}_{,j}$.
Taking the derivative of this expression with respect to $i$ gives
${T^{i0}}_{,0i}=-{T^{ij}}_{,ij}$.
Combining the two equations obtained, we find
${T^{00}}_{,00}={T^{ij}}_{,ij}$, i.e., $\frac{\partial^2
T^{00}}{\partial {x^0}^2}= \frac{\partial^2 T^{ij}}{\partial
x^i\partial x^j}$.
Let us multiply both sides of this equality by $x^m x^n$, with
$m,n=1,2,3$, and integrate over all space. The right-hand side of the
equation can be integrated by parts twice to yield $2\int
T^{mn}d^3x$.
Thus, setting the time coordinate $x^0=ct$, we obtain from the
equation above that $\frac{1}{c^2}\frac{\partial^2 }{\partial t^2}
\int T^{00} x^mx^nd^3x=2\int T^{mn}d^3 x$.
We then deduce that the metric perturbation from
Eq.~\eqref{habintegrado} is given by $h^{mn}(r,t)=\frac{2G}{c^6
r}\frac{\partial^2 }{\partial t^2} \int T^{00}x^mx^nd^3 x$.
Now, for low velocities of the particles in the source, we have
$T^{00}=\rho\, c^2$, where $\rho$ is the mass density of the matter
that acts as the source of gravitational waves. Hence, we can write
$\int T^{00}x^mx^nd^3 x=\int \rho\,c^2 x^mx^nd^3 x$.
But the integral $\int \rho\, x^mx^nd^3 x$ is the second moment of
inertia of the source, called the quadrupole moment of the mass
distribution, and denoted by $I^{mn}$. Thus, $I^{mn}=\int \rho
x^mx^nd^3 x$.
Therefore, from Eq.~\eqref{habintegrado} and the results above, we
obtain that the gravitational-wave perturbation is
\begin{equation}
h^{ij}(r,t)=\frac{2G}{c^4 r} \ddot{I}^{ij}\left( t_r\right),
\label{hijIij} 
\end{equation}
where a dot over a quantity denotes a derivative with respect to time,
we have replaced the symbols $m,n$ by $i,j$, and $t_r= t-\frac{r}{c}$
is the retarded time.
This is the famous quadrupole formula for gravitational radiation. It
states that the metric at a given position $r$ is given by the second
time derivative of the quadrupole moment of the mass distribution, and
is inversely proportional to the distance $r$ from the source,
evaluated at the retarded time. With it, we can calculate much more.

Let us now calculate the luminosity in gravitational energy that emanates
from the source.
We define the energy density in the gravitational field as
$\tau_{00}$.
From Einstein's equation, Eq.~\eqref{Eequation}, we have
$G_{00}=\frac{8\pi G}{c^4}\tau_{00}$.
Using $g_{ab}=\eta_{ab}+h_{ab}$, Eq.~\eqref{gwmetric}, one obtains after
some calculation an expression for  $\tau_{00}$, given by
$\tau_{00}=\frac{c^4}{32\pi G}
h_{ij,0}{h^{ij}}_{,0}$.
Substituting $h^{ij}$ by
$h^{ij}(r,t)=\frac{2G}{c^4 r} \ddot{I}^{ij}\left( t_r\right)$,
see Eq.~\eqref{hijIij},
we obtain that the gravitational energy has the form
$\tau_{00}= \frac{G}{8\pi
c^6r^2}\dddot{I}_{ij}\dddot{I}_{ij}$.
We wish to calculate the rate of energy loss through a spherical
surface centered on the source and with large radius.
The gravitational energy passing through a spherical surface per unit
time is by definition given by
$\frac{dE}{dt}=\int c \tau_{00} r^2d\Omega$
where $d\Omega$ is an element of solid angle at any $r$.  Using the
expression for $\tau_{00}$, we find that
$\frac{dE}{dt}$ is $\frac{dE}{dt}= \frac{G}{8\pi c^5}\int
\dddot{I}_{ij}\dddot{I}_{ij}d\Omega$.
This expression must be handled carefully, because only the traceless
and transverse part of $I_{ij}$ in the direction of energy flow
contributes.
Taking this into account and performing the angular integration, one
can, after some calculation, find the numerical coefficient that
multiplies the quantity
$\frac{G}{c^5}\dddot{I}_{ij}\dddot{I}_{ij}$.
This coefficient is $\frac{8\pi}{5}$,
and thus the gravitational luminosity of
the source $L\equiv\frac{dE}{dt}$ is given by
\begin{equation}                        
L=
\frac{G}{5 c^5}\dddot{I}_{ij}\dddot{I}_{ij}.
\label{Luminosity}
\end{equation}
Hence, the gravitational luminosity of the source depends on the
square of the third time derivative of the quadrupole moment of
inertia.
Let us examine the units of each term in Eq.~\eqref{Luminosity}.  The
term $\frac{G}{5 c^5}$ has units of inverse luminosity.
Thus, for the units on both sides to match, each term $\dddot{I}_{ij}$
must have units of luminosity, as can be easily verified directly.
It is clear from the equation that $\frac{G}{5 c^5}$
is an important quantity in
gravitational radiation.
Since it is an extremely small number, it means that in order to
obtain a detectable signal, the third time derivative of the
quadrupole moment of inertia,  $\dddot{I}_{ij}$, must be huge.
Everything that moves emits gravitational waves, but few sources
possess this property of having an enormous $\dddot{I}_{ij}$ term.

\subsection{A compact binary system of compact
objects as source of gravitational radiation}

We first need to calculate the quadrupole moment of inertia
$I_{ij}$ for a system that emits strong gravitational waves. As we have
seen, a compact binary system, i.e., a system
with a short orbital
separation composed of black holes or neutron stars, is such a system.
These objects are required because they concentrate a large amount of
mass within the small region defined by their orbit, allowing for the
efficient generation of gravitational waves.

\begin{figure*}[h]
\centering
\includegraphics[scale=0.6]{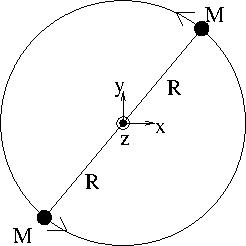}
\caption{
Two compact objects, each of mass $M$, in a tight circular orbit of radius
$R$. For the emission of gravitational waves intense enough to be
detected, the objects must be either black holes or neutron stars.
}
\label{binaryorbitpt}
\end{figure*}

Let us then suppose two compact objects in a tight orbit around their
common center of mass. Each of the orbiting objects has mass $M$, and
the orbital radius is $R$, see Fig.~\ref{binaryorbitpt}.

For the purposes of our calculation, we assume that Newtonian gravity
provides an adequate approximation at this stage of the binary's
evolution, allowing us to treat each component as a point mass. Under
this assumption, the angular velocity $\Omega$
of the orbit, which is also
the orbital angular frequency, is given by
\begin{equation}                        
\Omega=\left(\frac{GM}{4R^3}\right)^{\frac12}.
\label{orbitalangularvelocity}
\end{equation}         

We write $x^i=(x^1,x^2,x^3)=(x,y,z)$.
At time $t$, according to the figure, one of the stars is at the position
 $(x,y.z)=(R\cos\Omega t,R \sin\Omega t,0)$,
and the other at
 $(x,y.z)=(-R\cos\Omega t,-R\,\sin\Omega t,0)$.
We want to calculate the moment of inertia $I^{ij}$
of this two-mass
system with respect to the axis perpendicular to the orbital plane and
passing through the origin.
Since these are two point masses, the integral of
the quadrupole moment of inertia,
$I^{ij}=\int \rho\, x^ix^jd^3 x$,
reduces to a sum over the two masses, giving
${I}^{ij}=2M\,x^i(t)x^j(t)$.
Therefore, the components are
$I^{xx}=
2MR^2\cos^2\Omega t=MR^2(1+\cos2\Omega t)$, $I^{yy}=2MR^2\sin^2\Omega
t=MR^2(1-\cos2\Omega t)$, and $I^{xy}=I^{yx}=2MR^2\cos\Omega t\,
\sin\Omega t=MR^2\,\sin2\Omega t$,
where we have used trigonometric identities.
In matrix form, this becomes
\begin{equation}   
{I}^{ij}(t)=MR^2\begin{pmatrix}
1+\cos2\Omega t &\sin2\Omega t&0\\
\sin2\Omega t&1-\cos2\Omega t &0\\
0&0&0
\end{pmatrix},
\label{Iij1}
\end{equation}         
with $i,j=1,2,3$.
By differentiating this expression twice with respect to time, we obtain
\begin{equation}   
\ddot{I}^{ij}(t)=-4\Omega^2MR^2
\begin{pmatrix}
\cos2\Omega t &\sin2\Omega t&0\\
\sin2\Omega t&-\cos2\Omega t &0\\
0&0&0
\end{pmatrix},
\label{ddotIij1}
\end{equation}
where a dot over a quantity denotes differentiation with respect to
time.  Using Eq.~\eqref{ddotIij1}, we can write $h^{ij}$, from
Eq.~\eqref{hijIij}, as
\begin{equation}   
h^{ij}(t,r)=\left( \frac{2GM}{c^2R} \right)
\left(\frac{2GM}{c^2r}\right)
\begin{pmatrix}
\cos2\Omega t_r &\sin2\Omega t_r&0\\
\sin2\Omega t_r&-\cos2\Omega t_r &0\\
0&0&0
\end{pmatrix},
\label{hijduasmassas}
\end{equation}  
with $t_r= t-\frac{r}{c}$, the retarded time.
We see that Eq.~\eqref{hijduasmassas} has the same form as
Eq.~\eqref{hab+x}.
Therefore, it represents radiation in the  $z$ direction.
Note that by comparing the two equations we find that the amplitude $h$
of the wave is
 $h=\left( \frac{2GM}{c^2R} \right)
\left(\frac{2GM}{c^2r}\right)$,
and the frequency $\omega$ of the wave is $\omega=2\Omega$,
meaning that the
gravitational wave frequency is twice the orbital frequency.
The reason is clear, exchanging one of the objects with the other does
not change the configuration of the system, so each half orbit
corresponds to one oscillation of the wave, one full orbit
corresponds to two oscillations.
Let us recall that $R$ is the orbital radius and $r$ is the distance
from the orbit to the detector.
We know that  $\frac{2GM}{c^2}$
is the Schwarzschild radius, or the gravitational
radius, of each object, and if the object is a black hole, this
radius coincide with the its event horizon radius.
Let us insert some numbers. For example,
$\frac{2GM}{c^2 R}\simeq0.1$.
If $\frac{2GM}{c^2}\simeq100\,{\rm Km}$,
and $r\simeq 100\,{\rm Mpc}\,\simeq10^{21}\,{\rm Km}$,
we obtain $h\simeq 10^{-20}$, an extremely small value.
We thus see that only very massive objects in tight orbits can be
detected at large distances.

To obtain the gravitational luminosity from Eq.~\eqref{Luminosity}, we
need to calculate the third time derivative of the quadrupole moment
of inertia. From Eq.~\eqref{ddotIij1}, this is given by
\begin{equation}
\dddot{I}^{ij}(t)=8\Omega^3MR^2
\begin{pmatrix}
\sin2\Omega t &-\cos2\Omega t&0\\
-\cos2\Omega t&\sin2\Omega t &0\\
0&0&0
\end{pmatrix}.
\label{ddotIijbinary}
\end{equation}  
Now, $L\equiv \frac{dE}{dt}=
\frac{G}{5 c^5}\dddot{I}_{ij}\dddot{I}_{ij}$, see Eq.~\eqref{Luminosity}.
Performing the calculation for
$ \dddot{I}^{ij}\dddot{I}^{ij}=\dddot{I}^{11}\dddot{I}^{11}
+
2\dddot{I}^{12}\dddot{I}^{12}
+
\dddot{I}^{22}\dddot{I}^{22}$,
we find from Eq.~\eqref{ddotIijbinary} that
$\dddot{I}^{ij}\dddot{I}^{ij}=128 M^2R^4\Omega^6$.
We then obtain  $L=\frac{128G}{5 c^5}M^2R^4\Omega^6$.
But, since $\Omega=\left(\frac{GM}{4R^3}\right)^{\frac12}$,
see Eq.~\eqref{orbitalangularvelocity},
the gravitational luminosity becomes
\begin{equation}
L=\frac25\left(\frac{GM}{Rc^2}\right)^5\frac{c^5}{G}.
\label{Luminositybynary}
\end{equation}
This luminosity corresponds to the energy emitted along the rotation
axis $z$.  Radiation emitted in the $x$ or $y$ directions can also be
calculated.
We will not do so here, but it can be shown that the energy flux in
those perpendicular directions is eight times smaller than that
emitted along the rotation axis, and can be neglected as a first
approximation.

Now, the total energy $E$ of the binary system is given by $E=T+U$,
where $T$ is the kinetic energy of the system and $U$ its potential
energy.  We have $T=2\times\frac12 MR^2\Omega^2= \frac{GM^2}{4R}$,
where Eq.~\eqref{orbitalangularvelocity} has been used, and
$U=-\frac{GM^2}{2R}$. Hence,  $E=-\frac{GM^2}{4R}$.
The orbital period $P$ is
$P=\frac{2\pi}{\Omega} =4\pi
\left(\frac{R^3}{GM}\right)^{\frac12}$.
Then, $\frac{dP}{P}=\frac32
\frac{dR}{R}$. 
Since the system is losing energy, $R$ decreases.
Thus, as $\frac{dR}{R}=-\frac{dE}{E}$, we obtain
 $\frac{dP/dt}{P}=-\frac32
\frac{dE/dt}{E}=-\frac32 \frac{L}{E}$.
Using Eq.~\eqref{Luminositybynary}, we find
$\frac{dP/dt}{P}=
-\frac{12}{5}\frac{G^3M^3}{R^4c^5}$,
where the expression for  $E$ was also used.
Thus,
$\frac{dP}{dt}=-\frac{12}{5}\frac{G^3M^3}{R^4c^5}\,P$,
or, using  $P=4\pi \left(\frac{R^3}{GM}\right)^{\frac12}$,
we obtain
$\frac{dP}{dt}=-\frac{12}{5}(4\pi)^{\frac83}\left(\frac{GM}{c^3P}
\right)^{\frac53}$.
Inverting this equation for the mass $M$, we find
$M=\frac{c^3}{G}
\left(-\frac{5}{96}\pi^{-\frac83}2^{-\frac43}P^{\frac53}\dot P
\right)^{\frac35}$.
But, the detector directly measures the frequency $f$
of the gravitational wave.
Since  $f=2\Omega=\frac{2}{P}$, we obtain
$M=2^{\frac15}\hskip -0.1cm
\left(\frac{5}{96}\pi^{-\frac83}f^{-\frac{11}{3}} \dot f
\right)^{\frac35}\frac{c^3}{G}$.
This formula governs the initial inspiral phase
for two equal masses in orbit. Thus,
\begin{equation}
\frac{M}{2^{\frac15}}
=\left(\frac{5}{96}\pi^{-\frac83}f^{-\frac{11}{3}}
\dot f \right)^{\frac35}\frac{c^3}{G}.
\label{chirpequalmasses}
\end{equation}	
The chirp mass,
named after the characteristic rising chirp sound
of the signal, is defined for two objects of equal mass as
 $M_{\rm c}=\frac{M}{2^{\frac15}}$.
Hence, for two equal-mass objects in mutual orbit, we have
 $M_{\rm c}=
\left(\frac{5}{96}\pi^{-\frac83}f^{-\frac{11}{3}} \dot f
\right)^{\frac35}\frac{c^3}{G}$.
From Eq.~\eqref{chirpequalmasses}, we see that given $f$ and $\dot f$,
measured at the moment of gravitational wave detection, we can
determine $M$ or $2M$, the total mass of the two objects of equal
mass.  When the objects have different masses, the calculation follows
the same procedure and is not qualitatively more complicated.

All this derivation was carried out with the aim of determining the
mass of a binary system of compact objects in mutual orbit. Using the
calculations above for the emission of gravitational energy, one can
easily study the characteristics of the orbital radius decrease in
such binary systems, as was done for the famous binary pulsar, leading
to the conclusion that the system loses energy through the emission of
gravitational waves.

\vskip -1.1cm 
\centerline{}

\subsection{The three phases of the compact binary orbits
in gravitational wave emission}

\vskip -1.1cm 
\centerline{}

\begin{figure*}[h]
\centering
\includegraphics[scale=0.35]{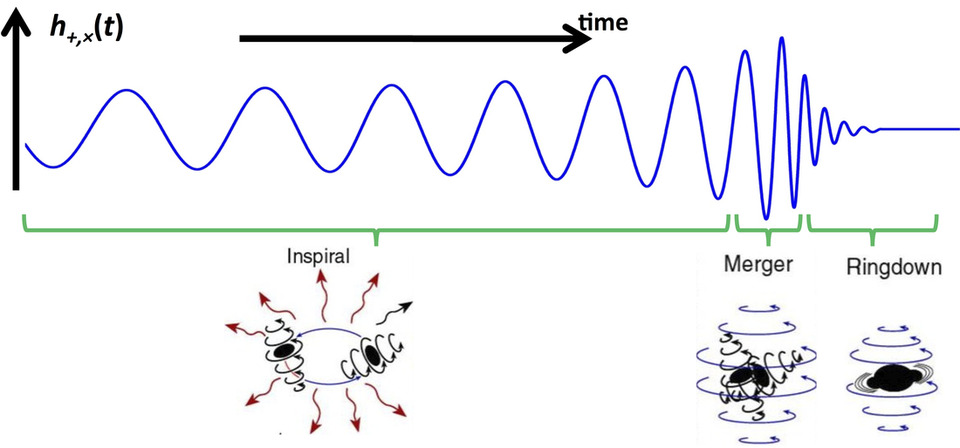}
\caption{
The three phases of the orbit in gravitational wave emission, shown in
the strain $h$
for both the $+$ or $\times$  polarizations as a function of
time.
}
\label{inspiralmergerringdown2}
\end{figure*}

There are three phases of the compact binary orbit in the emission of
gravitational waves,
see Fig.~\ref{inspiralmergerringdown2}.
The inspiral phase, which we studied above, can be calculated using a
Newtonian approximation, or more accurately, with a post-Newtonian
approximation using methods of the type we employed here. The
frequency has the shape of a chirp, a rising tone  that increases
with time.
The merger phase which can only be calculated through powerful numerical
methods using supercomputers.
The ringdown phase, in which the final system settles down,
requires perturbation theory in general relativity. The theory
provides the
quasinormal modes of vibration of the final black hole resulting from
the merger.

\subsection{Main properties of gravitational waves}

\noindent
The main properties of gravitational waves can be summarized as follows:

\noindent
$\cdot$They propagate at the speed of light, $c$.

\noindent
$\cdot$They are transverse. Waves propagating in the
$z$-direction obey $h_{zz}=0$.

\noindent
$\cdot$They have two polarizations: the $+$ polarization and the
$\times$ polarization.

\noindent
$\cdot$They are tidal forces, i.e., they possess nonzero curvature
and change the distance between

\hskip -0.3cm test masses.

\noindent
$\cdot$They carry energy.

\noindent
$\cdot$They are generated by any mass, or system of masses, undergoing
    quadrupolar vibration,

\hskip -0.3cm 
in particular, by a binary system of
    compact stars.

\noindent
$\cdot$They have properties analogous to electromagnetic \hskip -0.05cm
waves:
\hskip -0.05cm    propagation speed, transversality,

\hskip -0.3cm 
polarization, and an amplitude
    that decreases inversely with distance. Note that this

\hskip -0.3cm 
decay
    implies that doubling the detector's sensitivity increases the
    observable volume of 

\hskip -0.3cm 
the universe by a factor of eight.

\noindent
$\cdot$They also share some properties with sound: difficulty of being
    absorbed; the emitted

\hskip -0.3cm 
wavelength is comparable to the size of the
    source, which makes imaging difficult. It is

\hskip -0.3cm 
worth noting that the
    frequencies of events involving stellar-mass black holes or
    neutron

\hskip -0.3cm 
stars lie within the range audible to the human
    ear. However, for these signals to be

\hskip -0.3cm 
effectively heard, an
astronomical amplification is required.

\section{New Cathedrals of Science}

\begin{figure*}[h]
\centering
\includegraphics[scale=0.7]{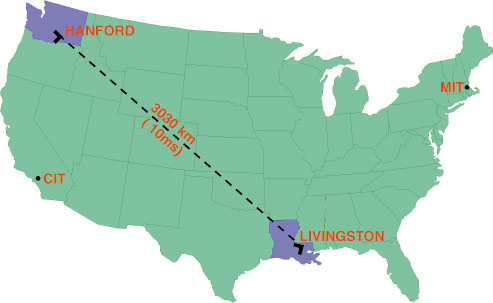}
\caption{The locations of the two LIGO detectors,
one in Hanford, Washington, and the other in Livingston, Louisiana,
as well as the Caltech, also known as CIT, and MIT
research centers.
}
\label{doisdetetores}
\end{figure*}

\begin{figure*}[h]
\centering
\includegraphics[scale=0.131]{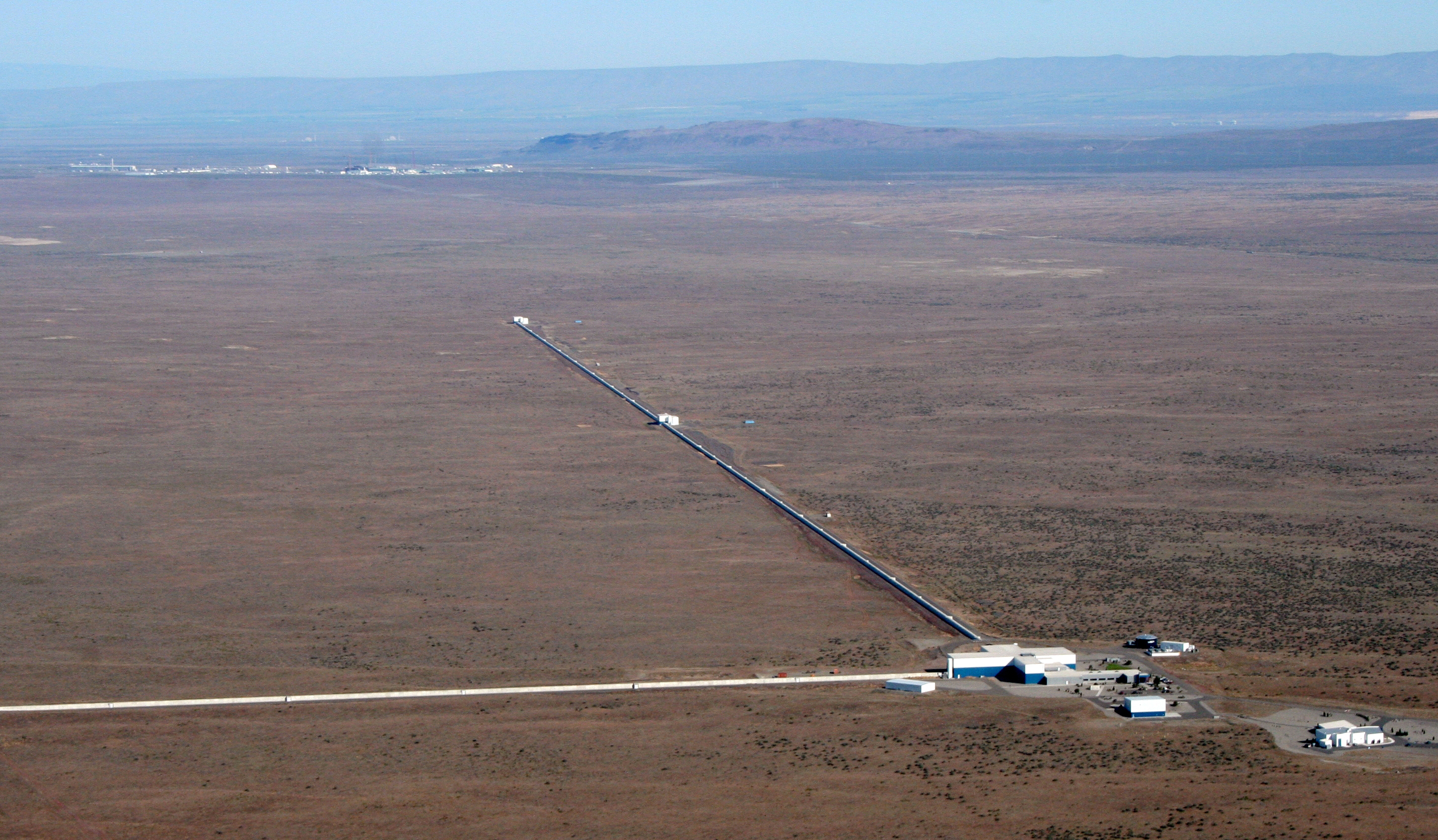}
\caption{View of the LIGO observatory in Hanford, Washington.
The two 4 Km arms forming an L shape are clearly visible.
}
\label{hanfordligo}
\end{figure*}

\begin{figure*}[h]
\centering
\includegraphics[scale=1.65]{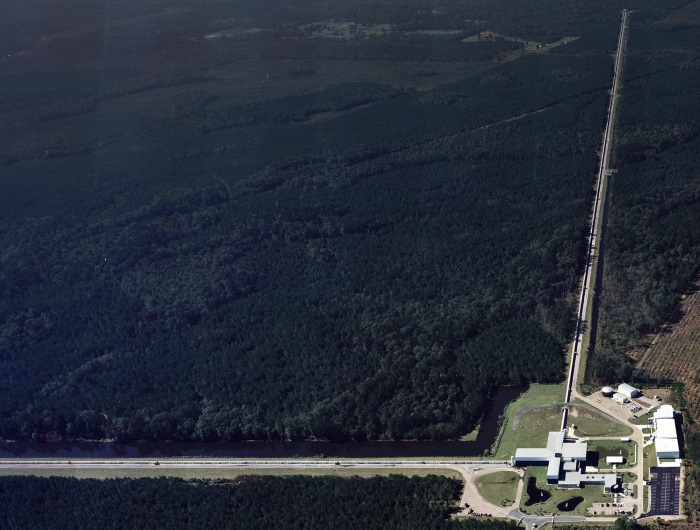}
\caption{Partial view of the LIGO observatory in Livingston, Louisiana.
This site also employs arms with 4 Km.
}
\label{livingstonligo}
\end{figure*}

Two new cathedrals of science arose just after the year 2000: the
two LIGO detectors. These structures were designed to explore the
universe through gravitational waves. The LIGO detectors are located
in the United States, in two distinct regions chosen to maximize the
precision of the measurements. One is in Hanford, in the state of
Washington, in the far northwest of the country, and the other is in
Livingston, Louisiana, in the south-central region, see
Figs.~\ref{doisdetetores}, \ref{hanfordligo}, and
\ref{livingstonligo}.
The distance between the two, about 3000 Km, is important for
confirming the detection of gravitational waves and for eliminating
local interferences. This separation makes it possible to verify
whether a detected signal is indeed a gravitational wave, since the
signal will reach both detectors with a small time delay on the order
of milliseconds, or merely local noise, which would affect only one
of the detectors. The distance between the two sites also helps
estimate the direction in the sky from which the wave originated.
The two LIGO detectors are the result of a joint initiative between
two of the most prestigious scientific institutions in the United
States: the Massachusetts Institute of Technology (MIT) in Boston and
the California Institute of Technology (Caltech or CIT) in Pasadena.
The project secured funding from the National
Science Foundation in 1980, but it was only in 1994 that the
construction of the experimental facilities began, marking the start
of this remarkable scientific endeavor.

\begin{figure*}[h]
\centering
\includegraphics[scale=1.7]{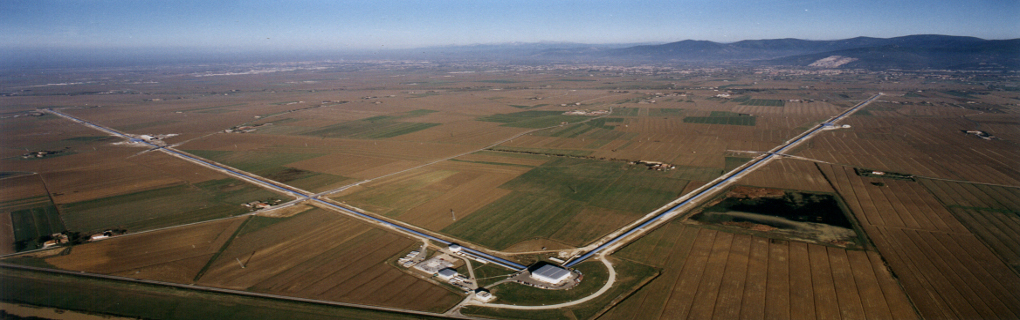}
\caption{The Virgo Observatory in Santo Stefano a Macerata, near Pisa,
Italy. It is part of the European Gravitational Observatory. Viaducts
were built above each of the 3 Km arms to allow vehicles from
neighboring farms to pass.
}
\label{virgo}
\end{figure*}

\begin{figure*}[h]
\centering
\includegraphics[scale=0.33]{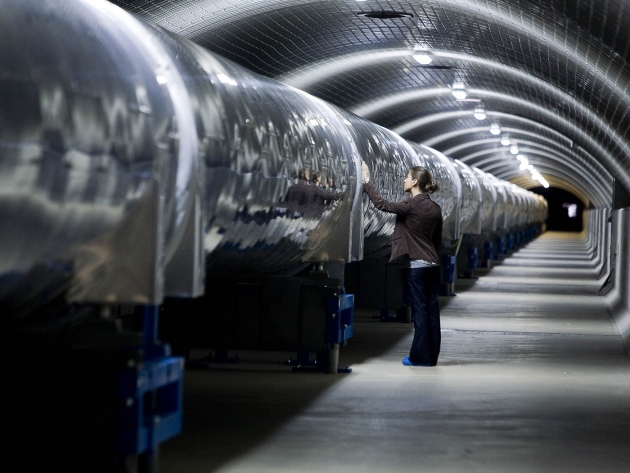}
\caption{Gigantic vacuum tube housing the lasers along the 3
km arms of the Virgo Observatory.
}
\label{virgopisainside}
\end{figure*}

\vskip 0.5cm
Another cathedral of science, also designed to detect gravitational
waves, is the Virgo observatory, located near Pisa, Italy, see
Figs.~\ref{virgo} and \ref{virgopisainside}. The initiative started in
the 1990s as a collaboration between  Italy
and France, with financial
support from the CNRS, i.e., the
Centre National de la Recherche Scientifique,
and the INFN, i.e.,
the Istituto Nazionale di Fisica Nucleare. The detector was
completed in 2005. Today, Virgo is operated and maintained by the EGO,
i.e., the European Gravitational Observatory, an international organization
headquartered at the site of the detector itself.


\vskip 3cm

Yet another gravitational wave detector is KAGRA, an underground
gravitational wave observatory located in Japan, which began
operations in 2023, and has arms of 3 Km.
LIGO, Virgo, and KAGRA operate jointly in a
collaboration known as LIGO-Virgo-KAGRA collaboration.

\begin{figure*}[h]
\centering
\includegraphics[scale=0.6]{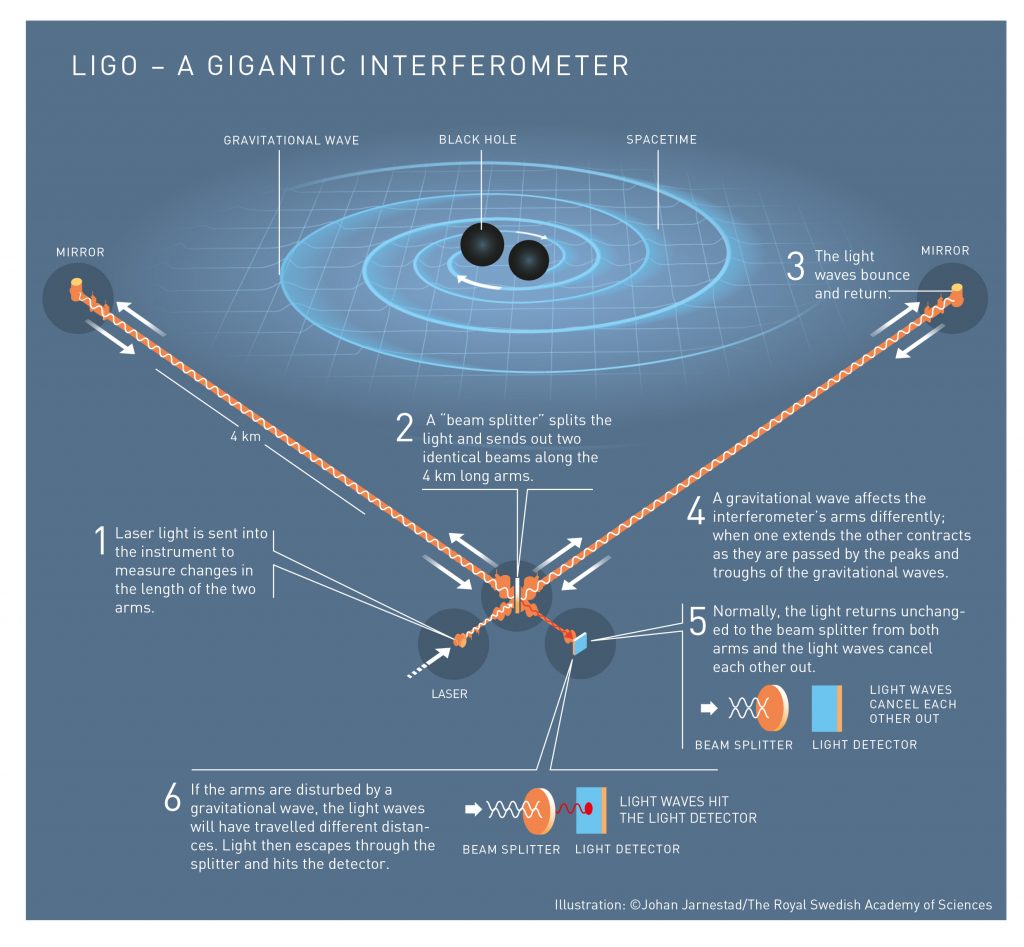}
\caption{Scheme of the LIGO interferometer with 4 Km arms, a giant
interferometer. This diagram also applies to the other operational
interferometers, such as Virgo and KAGRA which have
arms of 3 Km. See the figure and main text
for details. The figure is taken from an illustration by Jarnestad
of the Royal Swedish Academy of Sciences.
}
\label{ligosketch}
\end{figure*}

Gravitational-wave observatories such as LIGO, Virgo, and KAGRA employ
large-scale Michelson interferometers,
see Fig.~\ref{ligosketch} for details.
 They operate
state-of-the-art technology in optics, seismic isolation, and noise
control to achieve unprecedented sensitivity, being capable of
measuring relative length variations on the order of
$h=\frac{\Delta L}{L}=10^{-21}$,
which corresponds to distortions in spacetime smaller than
one-thousandth of a proton's diameter along their interferometric
arms. Together, the four instruments
significantly enhance the ability to detect gravitational waves
reaching Earth with extremely small amplitudes, originating from the
most energetic events in the universe, such as mergers of black holes
or neutron stars at cosmological distances.

\begin{figure*}[h]
\centering
\includegraphics[scale=0.4]{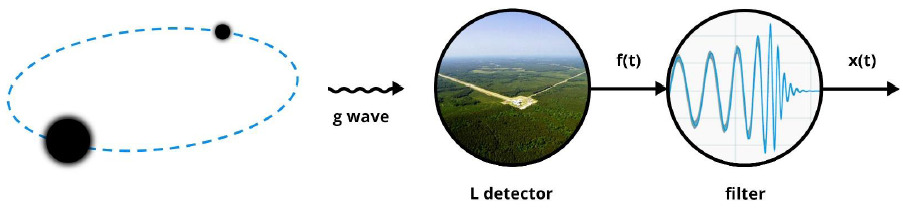}
\centerline{(a)}
\includegraphics[scale=0.6]{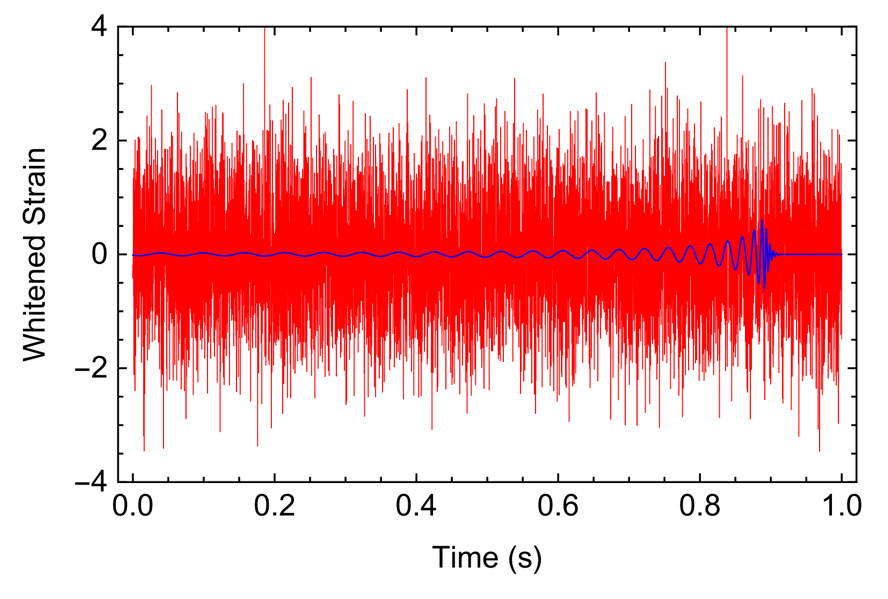 }
\centerline{(b)}
\caption{(a) Diagram of the entire process, from the generated wave,
to the wave detected by the L-shaped detector, the filtered wave, and
finally the strain.  (b) Visualization of the signal filtering
technique by waveform matching, i.e.,
matched filtering technique. The
signal is completely buried in noise, but by matching it with modeled
gravitational wave signals, the desired information can be extracted.
}
\label{matchfiltering}
\end{figure*}

These detectors record mostly noise, such as seismic, thermal,
electronic, and quantum noise, see Fig.~\ref{matchfiltering}.
Gravitational wave signals are embedded within this background
noise. Extracting these signals from the noise is a major challenge,
requiring sophisticated data analysis methods. One of the most
effective techniques in this process is the matching of theoretical
waveform templates with observational data, known as the matched
filtering technique. This method consists of correlating the collected
data with a library of waveform templates previously computed from
numerical models of general relativity, allowing the identification of
real signals even when their amplitude is extremely low. This
technique has proven highly successful and was of decisive importance
for the detection of the first gravitational wave events.

To locate the source of a gravitational wave, temporal triangulation
is used. This procedure consists of constructing possible triangles from
the source to the existing detectors, combined with the analysis of
the differences in arrival times of the waves at each detector. Since
gravitational waves travel at the speed of light, these arrival-time
differences, even of just a few milliseconds, can be sufficient to
determine the direction of the source. Clearly, the time delay between
detections indicates that the gravitational wave comes roughly from
the direction of the detector that registered it first.
With only one detector, triangulation is impossible, and the source
cannot be located. With two detectors, temporal triangulation based on
their locations allows the source to be constrained to a ring, a halo
in the sky, around the direction of the earliest detection. A third
detector reduces this ring to two antipodal points, which correspond
to the intersections of the two rings defined by each pair of
detectors. A fourth detector breaks this degeneracy completely,
providing the exact point in the sky where the event occurred.
Of course, the detectors must be far apart, four detectors in the
same place, or very close to each other, would be of no
use. Therefore, having several detectors distributed across the planet
is essential to localize the source with reasonable accuracy. This is
already achieved with the two LIGO detectors, along with the Virgo and
KAGRA detectors.

\section{The historical event: The GW150914 and its source}

\begin{figure*}[h]
\centering
\includegraphics[scale=0.5]{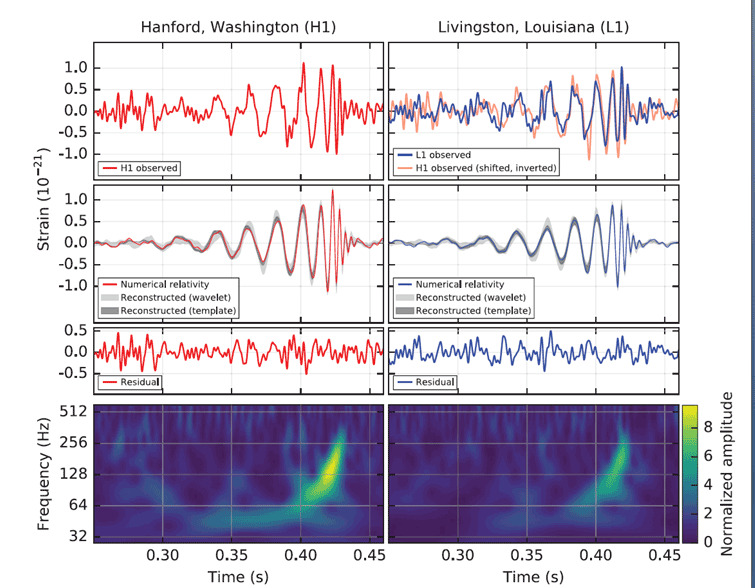}
\caption{
Plots of the signals from the first detected gravitational wave, as
published by the LIGO Collaboration in the American journal Physical
Review Letters on February 12, 2016. See text for details.
}
\label{thehistoricaleventGW150914}
\end{figure*}

The first detection of a gravitational wave by humanity occurred on
September 14, 2015. The signal was recorded almost simultaneously by
the two operational LIGO interferometers, located in Hanford and
Livingston, with the latter detecting it about 7 milliseconds earlier,
at 4:50:45 am USA Central Daylight
Time. After rigorous
verification and data analysis, the LIGO collaboration confirmed the
authenticity of the event, designated GW150914, as originating from
the merger of two black holes with masses of approximately 29 and 36
times that of the Sun. The results of this discovery were
published in Physical Review Letters on February 12, 2016, marking the
dawn of gravitational-wave astronomy and confirming one of the
fundamental predictions of Einstein's general theory of relativity,
see Fig.~\ref{thehistoricaleventGW150914}.

Let us examine  Fig.~\ref{thehistoricaleventGW150914}.
On the left is the Hanford signal, and on the
right, the Livingston signal. The top three rows of panels correspond
to the strain $h$ as a function of time. The first top row shows the
signal detected by each interferometer. In the right-hand panel, the
Hanford signal is superimposed on the Livingston one to demonstrate
that both correspond to the same wave.
The second row shows the
numerical template signal, on both the left and right, that best fits
the detected signal shown in the row above. From the figure, we can see that at
0.35 s the relative strain amplitude is about 0.25, while at 0.42 s it
reaches about 1, four times larger. Both values must be multiplied by
$10^{-21}$. The third row shows the residual signal between the first
and second rows, which in our context is of little further
interest. The fourth row, in blue, displays the frequency of the
signal as a function of time. From 0.30 s to 0.42 s, the frequency
increases, over about six cycles, from 35 Hz to 250 Hz, the point at
which the amplitude reaches its maximum.
The explanation for this evolution, i.e., increasing amplitude and
frequency, is the inspiral phase, caused by the emission of
gravitational waves from two orbiting masses, $M_1$ and $M_2$, that
eventually merge at about 0.43 s. The figure clearly shows the
ringdown phase that follows the merger. Thus, the three characteristic
phases of a compact binary system are all visible in the figure
representing the first-ever detection of a gravitational wave.

The initial evolution when the frequencies are low is
characterized by
the chirp mass $M_{\rm c}$.
As we have seen, if the masses of the two objects
are equal, $M_1=M$ and $M_2=M$, the chirp mass is defined as
$M_{\rm c} \equiv  \frac{M}{2^{\frac15}}$.
For masses $M_1$ and $M_2$
that differ from each other, the calculation
is a bit more involved than what we have done before. The chirp mass
is defined as
\begin{equation}
M_{\rm c}
\equiv \frac{(M_1M_2)^{\frac35}}{(M_1+M_2)^{\frac15}}.
\label{chirpmassgeneral}
\end{equation}	
The calculations yield that the relationship between the chirp mass
$M_c$ and the frequency $f$ is
\begin{equation}
M_{\rm c}=\left(\frac{5}{96}\pi^{-\frac83}f^{-\frac{11}{3}}
\dot f \right)^{\frac35}\frac{c^3}{G}.
\label{freqchirpmassgeneral}
\end{equation}	
When $M_1=M_2\equiv M$, Eq.~\eqref{chirpmassgeneral} gives
$M_{\rm c}=\frac{M}{2^{\frac15}}$,
and Eq.~\eqref{freqchirpmassgeneral} yields
Eq.~\eqref{chirpequalmasses}.

We can now look carefully at the plots. From them, one can estimate
$f$ and $\dot f$. By inspecting the plot, we can extract an average
frequency of approximately $f=120\, {\rm Hz}$, and a value of $\dot f$
given by $\dot f=\frac{128-64}{0.03}=2130\,{\rm Hz}/{\rm s}$,
approximately.
Using Eq.~\eqref{freqchirpmassgeneral} with the known values of $G$
and $c$, we obtain $M_{\rm c}=22 M_\odot$, approximately. This was
only an estimate. The precise result gives $M_{\rm c}=30 M_\odot$.
We can make several deductions from Eq.~\eqref{chirpmassgeneral}. The
minimum chirp mass $M_c$ occurs when $M_1=M_2\equiv M$. In this case,
since the mass $M$ and the chirp mass $M_{\rm c}$ are related by
$M=2^\frac15M_c$, as saw above, we have $M=35 M_\odot$, so
that the total mass would be $70 M_\odot$.
Because the minimum occurs when the masses are equal, we can generally
write for this event that $M_1+M_2\geq 70M_\odot$.  The Schwarzschild
radius for the binary system then satisfies
$\frac{2G(M_1+M_2)}{c^2}\geq210\, {\rm Km}$.
We can make further deductions from these data. A binary system of two
neutron stars has a maximum total mass of around $4M_\odot$, so it
does not have the required mass and can be excluded. A binary system
composed of a black hole and a neutron star with $M_{\rm c}=35
M_\odot$ would, according to the definition of $M_{\rm c}$,
Eq.~\eqref{freqchirpmassgeneral}, require the black hole to have a
very large mass, making the total mass of the system extremely
high. Thus, this configuration is also excluded a priori.
Therefore, what remains is a binary system of two black holes.
Moreover, a gravitational wave frequency of, say, $f=150\,{\rm Hz}$
means, as we have seen, that the orbital frequency is half of that,
$\Omega=75\,{\rm Hz}$. This implies that the two black holes are very
close, moving at velocities close to the speed of light, say, around
$\frac{c}{2}$.
Hence, the distance between them during this inspiral phase, the stage
we are considering, is of the order of $\frac{c}{2\Omega}=2000\,{\rm
Km}$. Each object has an extension of about
$200\,{\rm Km}$, and they are separated by $2000\,
{\rm Km}$.  This is a a lot of gravitation.
We are, in fact, dealing with compact objects in a tight
orbit.

Much more precise results can be achieved through numerical relativity
combined with the construction of computational templates adapted to a
large number of possible black hole collision events. From
Fig.~\ref{thehistoricaleventGW150914}, numerical relativity allows one
to extract precise numbers.
In Fig.~\ref{thehistoricaleventGW150914}, the three phases are shown
as perceived through the way the strain $h$ evolves over time. Thus,
for this first event, the models provide:

\noindent
$\cdot$Mass of the primary black hole: $M_1= 36 M_\odot$.

\noindent
$\cdot$Mass of the secondary black hole: $M_2= 29 M_\odot$.

\noindent
$\cdot$Mass radiated in gravitational waves: $M_{\rm og}= 3 M_\odot$.

\noindent
$\cdot$Mass of the final black hole: $M_{\rm f}= 62 M_\odot$.

\noindent
$\cdot$Angular momentum of the final black hole:: $J_f= 0.67$.
 
\noindent
$\cdot$Distance to the event:  $410 \,{\rm Mpc}$, $z=0.1$.    

\noindent
Some notes are important here.  $M_\odot=2\times10^{30}\,{\rm
Kg}$ is the mass of
the Sun. The mass of the gravitational waves $M_{\rm og}$ represents the
mass equivalent of the energy radiated as gravitational waves. The
quantity $M_{\rm f}$ is the final mass of the resulting black hole, which
must clearly be smaller than the sum of the initial black hole
masses. The quantity   $J_f$, the final angular momentum of the resulting
black hole, is a measure of its rotation, which can also be inferred
from theory and observational data.
${\rm Mpc}$ means megaparsec, with $1\,{\rm pc}$  equal to  $3.3$
light-years. The
redshift $z$ is a distance indicator through Hubble's law.

The detection of the event GW150914 marked a historic moment in
astrophysics and gravitation. Several general implications can be
outlined:

\noindent
1. With this event, the existence of stellar-mass black holes with masses
greater than  $25 M_\odot$, i.e., 25 times the mass of the Sun, was
confirmed for the first time.

\noindent
2. 
The event demonstrated that binary systems composed of black holes do
in fact exist in nature and can merge within the age of the universe,
the Hubble time. Such systems had already been predicted
theoretically, especially in extremely dense astrophysical
environments such as star clusters, where dynamical gravitational
interactions favor the formation of black hole pairs. The confirmation
of this prediction was of great importance.

\noindent
3. 
The origin of such massive stellar black holes is linked to the
evolution of very massive stars in low-metallicity regions, i.e.,
environments poor in elements heavier than hydrogen and helium. This
is because in regions with high metallicity, stellar winds are intense
and tend to expel large amounts of mass from the star before its final
collapse, making it difficult for high-mass black holes to form.

\noindent
4. 
Analyses supported by this event confirm that the merger rate of
binary black holes is on the order of tens per cubic gigaparsec per
year, more precisely, about $30 \,{\rm Gpc}^{-3}{\rm yr}^{-1}$,
where $\rm Gpc$ means gigaparsec. This indicates
that such events are not rare in the universe and that there is still
much to learn about the underlying processes.

\section{The 2017 Nobel Prize in Physics}

After the historic announcement in February 2016 of the event that
took place on September 14, 2015, the first-ever detection of
gravitational waves, the scientific community immediately recognized
the unparalleled importance of the discovery.  Expectations quickly
formed that the Nobel Prize in Physics would eventually be awarded to
the central figures of the LIGO project, given the rule that no more
than three people may receive the prize. It was known, however, that
the 2016 prize had already been decided, since the selection process
normally concludes in January or early February each year. Thus, the
scientific community was convinced that the 2017 Nobel Prize would
almost certainly be conceded to gravitational waves. And that is
exactly what happened.

In October 2017, the Nobel Committee's stated: The Royal Swedish
Academy of Sciences has decided to award the 2017 Nobel Prize in
Physics, with half going to Rainer Weiss (LIGO/Virgo Collaboration)
and the other half to Barry C. Barish (LIGO/Virgo Collaboration) and
Kip S. Thorne (LIGO/Virgo Collaboration) for their decisive
contributions to the LIGO detector and the observation of
gravitational waves.

\begin{figure*}[h]
\centering
(a)
\includegraphics[scale=0.3]{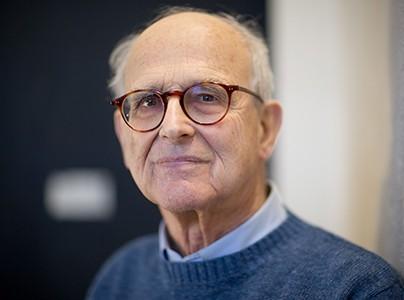}
\vskip 0.1cm
(b)
\includegraphics[scale=0.5]{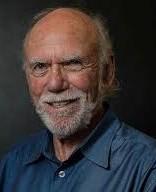}
\includegraphics[scale=0.085]{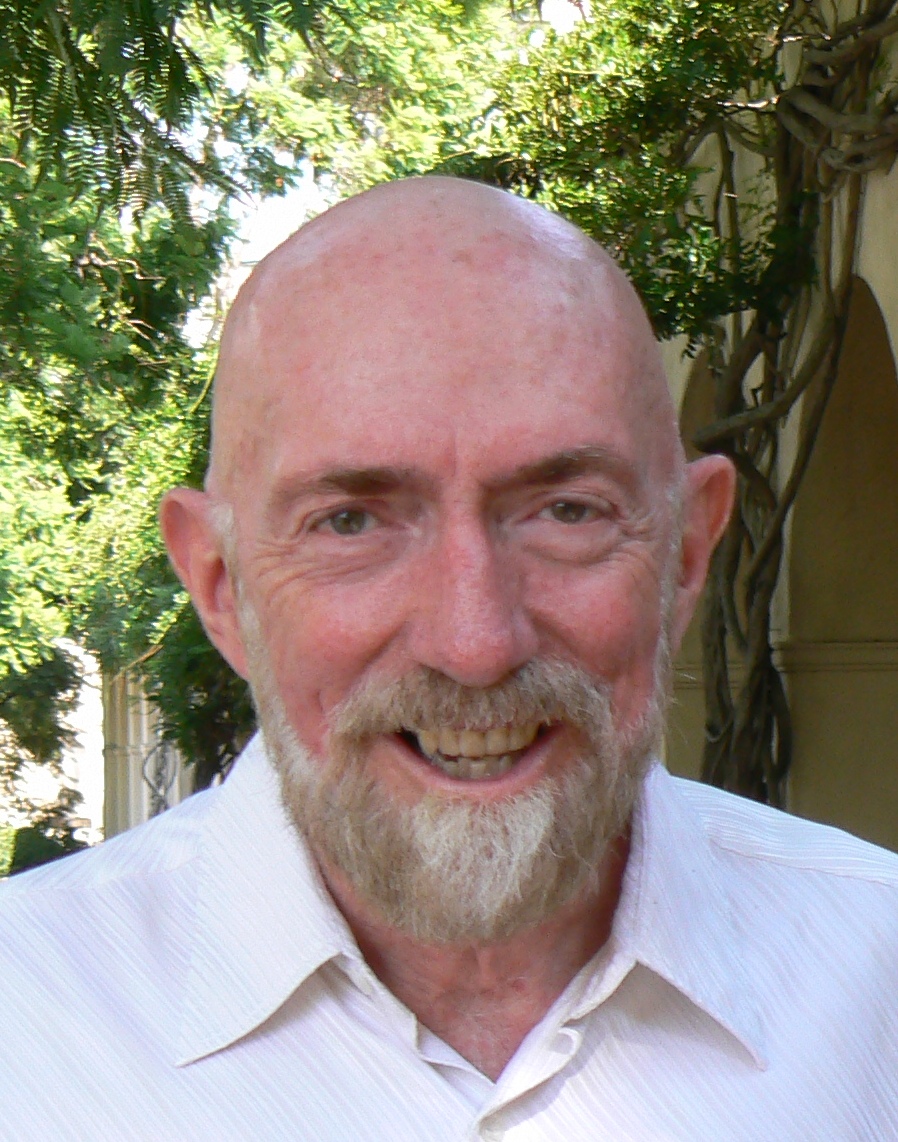}
\vskip 0.1cm
(c)
\includegraphics[scale=0.4]{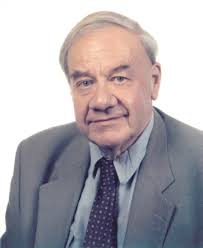}
\caption{(a) Rainer Weiss. (b) Barry Barish (left),
Kip Thorne (right). (c) Ron Drever.
}
\label{rainerweissetalportraits}
\end{figure*}

The distinction given to these three scientists was fully
deserved, see Fig.~\ref{rainerweissetalportraits}.
Weiss, affiliated with MIT and LIGO, received the first
half of the 2017 Nobel Prize in Physics for having invented the
appropriate detector for gravitational waves, a giant Michelson
interferometer.
Barish, affiliated with LIGO, and Thorne,
affiliated with Caltech and LIGO, received the second half of the
2017 Nobel Prize in Physics, the former for having led the LIGO
collaboration forward, and the latter for having developed, in a
systematic  way, the entire theory of gravitational
waves.
Drever,  the other
pioneering scientist in the development of LIGO and
equally worthy of the Nobel Prize, did
not live to see this recognition,
Fig.~\ref{rainerweissetalportraits}.

All the praise directed at the LIGO collaboration is insufficient. It
represents the highest example of how science should be done. Since
early on, the theory of general relativity predicted the existence of
gravitational waves. To confirm this prediction and open the way to a
new method of observing the universe, it was necessary to develop
detectors of virtually unimaginable precision. This achievement was
accomplished not only by the three scientists awarded the Nobel Prize,
but also by the hundreds of physicists, engineers, and specialists who
made up the LIGO collaboration. More than a mere scientific
experiment, the LIGO project represented an extraordinary milestone in
humanity's ability to probe the universe.

With the awarding of the 2017 Nobel Prize in Physics for the detection
of gravitational waves, generated by the collision of two black holes,
attention naturally turned to the black holes themselves, the
objects of gravitation par excellence, until then inaccessible to
direct observation.
Soon, many anticipated that a future Nobel Prize would recognize the
researchers whose insights and observations had transformed black
holes from speculation into established physics and astrophysics.
This anticipation
was fulfilled in 2020, when the Nobel Prize in Physics was granted to
three figures: Penrose, who mathematically demonstrated that black
holes are an inevitable consequence of general relativity, and Genzel
and Ghez, who independently led pioneering observations that
definitively confirmed the existence of a supermassive black hole at
the center of our galaxy. Thus, in the blink of an eye, in just a few
years, gravitational waves and black holes went from being almost
abstract theoretical predictions to becoming acknowledged protagonists
of contemporary physics.

\section{Other events and their sources}

The beginning of the era of gravitational astronomy came with the
detection of the first gravitational-wave signal by the LIGO
observatory, the famous event GW150914, produced by the merger of two
black holes that we have already discussed. This milestone opened a
new way of observing the universe through spacetime vibrations,
invisible to light. Since then, the LIGO detectors in the United
States, Virgo in Europe, and, more recently, KAGRA in Japan, have been
carrying out successive observation campaigns, designated by Oi, where
i represents the campaign number. The fourth campaign, O4, produced a
large number of detections. In total, these collaborations have
already identified more than 300 gravitational-wave events, including
confirmed detections and candidate events. The great majority of these
events correspond to mergers of black holes or neutron stars, although
some also involve mixed systems, such as a collision between a neutron
star and a black hole. The LIGO-Virgo-KAGRA collaboration continues to
analyze the data to identify new events and deepen our understanding
of the origin, frequency, and characteristics of these mergers. As
detector sensitivity increases, gravitational astronomy is expected to
reveal more details of these processes.

One of the most spectacular events detected by LIGO and Virgo occurred
during the O2 observation campaign. It was the event GW170817,
registered on 17
August 2017, which marked the first detection of gravitational waves
originating from the collision of two neutron stars. The parameters of
the collision are $M_1= 1.6\,M_\odot$, $M_2= 1.1
\, M_\odot$, $M_{\rm og}= 0.025 \, M_\odot$, $M_{\rm f}= 2.8 \,
M_\odot$, event distance $40\,{\rm Mpc}$.
GW170817 had extraordinary scientific importance because
it was the first event observed simultaneously through gravitational
waves and a broad range of electromagnetic signals, from gamma rays to
radio waves. The gamma-ray burst was detected about two seconds after
the arrival of the gravitational waves, in the galaxy NGC 4993,
approximately 130 million light-years from Earth,
i.e., $40\,{\rm
Mpc}$. This coincidence confirmed the link between
neutron-star mergers and the kilonova phenomena,
which are extremely
luminous explosions. It is now known that these explosions give rise
to the creation of heavy elements such as gold and platinum. In
addition, the Chandra X-ray Observatory detected, nine days later, an
emission coming from the event, suggesting the presence of an
expanding relativistic jet. Because X-ray emission is frequently
associated with jets formed intensely and with delay in rotating black
holes, some studies proposed that the final object of the merger was a
black hole. The event was also observed in optical, infrared, and
radio emission. Although the LIGO and Virgo detectors were able to
clearly capture the gravitational signal, their ability to accurately
locate the source of the event was limited at the time. Thanks to the
collaboration between dozens of observatories on Earth and in space,
it was possible to pinpoint the host galaxy. This event marks the
birth of multimessenger astronomy and enabled major advances in our
understanding of the origin of heavy elements such as gold, and in
measuring the expansion rate of the universe.

During the O3 observation campaign, in 2020, it was possible to
observe for the first time a collision between a black hole and a
neutron star, a type of event predicted by theory but not previously
detected. In the same year, the LIGO-Virgo collaboration identified
yet another event of this hybrid type, in which a black hole engulfs
its neutron-star companion. Another event of this kind was observed in
2023, during the O4 campaign. These events are particularly
significant because they combine the extreme characteristics of two of
the most compact known objects, namely,
the intense gravitational field of a
black hole and the ultra-dense matter of a neutron star. Observing
mergers between these two objects provides clues about the nature of
matter under extreme conditions and about the formation of binary
systems. Moreover, the detection of these signals helps complete the
catalog of possible compact-object mergers. Indeed, these detections
completed the trio of compact-star mergers in binary systems: black
hole-black hole, neutron star-neutron star, and black hole-neutron
star.

In 2025, the most massive black hole merger ever detected was
announced. The event, designated GW231123, was recorded in 2023 during
the fourth observation campaign O4 of the LIGO-Virgo-KAGRA
collaboration. The two black holes that merged have masses of
approximately $137\,M_\odot$ and $103\,M_\odot$.
The merger resulted in a
single black hole of about $225\,M_\odot$, setting a new record in
gravitational-wave detections. Besides their enormous mass, both black
holes had spins close to the limit allowed by general relativity. This
event poses a challenge to conventional stellar-evolution models,
which do not predict the formation of stellar-origin black holes with
such high masses, due to mass loss or even the explosion of the star
itself caused by the production of electron-positron pairs at its
core. This pair-instability occurs in stars with masses between 60 and
250 solar masses, and a gap in black hole masses would therefore be
expected in this range,
but this is not what is observed. Several
hypotheses exist.
One is that stellar models for high-mass, rapidly
rotating stars must be improved in order to explain the formation of
such black holes by gravitational collapse. Thus, this detection not
only represents a record in terms of mass, but also provides clues
about the existence and formation of intermediate-mass black holes,
with masses from  $200\,M_\odot$ to $10000\,M_\odot$, a class until
recently known only in theory.

Also in 2025, the event GW250114 of
a merger of two black holes with masses
of about $30\,M_\odot$
and low spins, showed that the area of the final
black hole is greater than the sum of the areas of the initial black
holes, thereby confirming Hawking's area theorem, also known as the
second law of black hole mechanics.

In Fig. 16 it is shown
the masses of black holes and neutron stars observed
both in the gravitational spectrum and in the electromagnetic
spectrum.

\begin{figure*}[h]
\centering
\includegraphics[scale=0.50]{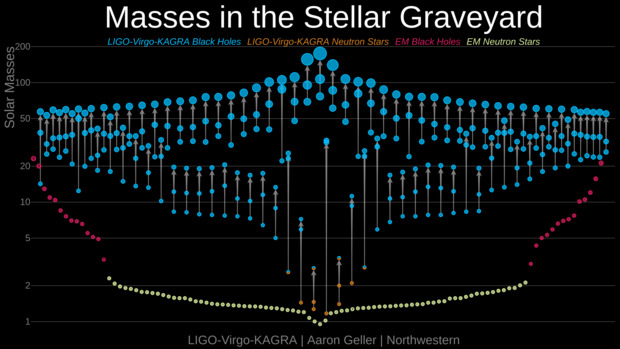}
\caption{
Masses of stellar black holes and neutron stars detected and
observed. The masses detected through gravitational waves by the LIGO
and Virgo collaborations are shown in blue for black holes and in
orange for neutron stars. The white arrows link the masses of the two
initial objects to the mass of the final object. In red and yellow
are shown the masses of black holes and neutron stars, respectively,
observed through electromagnetic processes. The numbers on the
vertical axis are given in units of solar masses. The figure is
taken from a figure by Elavsky and Geller of the LIGO-Virgo-Kagra
collaboration.
}
\label{graveyardstellargraveyardrunO3}
\end{figure*}

Another interesting aspect is that when the merger of two compact
objects occurs and the emission of gravitational waves is detected,
this can be used as a standard siren. It can be shown, though we will
not do so here, that the shape of the gravitational-wave signal makes
it possible to calculate directly the distance to the source. If, at
the same time, it is possible to identify the host galaxy and measure
its redshift $z$, one obtains a direct relation between distance $d$ and
redshift, exactly what is needed to estimate the Hubble constant $H_0$,
with the formula given by $d=H_0 z c$, where $c$ is the speed of
light. This method was successfully applied to the event involving the
collision of two neutron stars, the 2017 event GW170817. This
collision provided the first measurement of the Hubble constant using
gravitational waves, giving a value $H_0=70\,{\rm
Km/s/Mpc}$.  This
achievement inaugurated an entirely independent approach in cosmology
for determining the Hubble constant, one that is based on
gravitational waves rather than electromagnetic radiation. The value
obtained by the gravitational-wave method lies within the range of the
values obtained by other electromagnetic and cosmological methods.

\section{Cosmological gravitational waves and the Big Bang}

The gravitational waves that can be measured with current or
near-future technology are not only produced by black hole or
neutron-star mergers. The theory predicts that the primordial
universe, in its earliest moments, was an extremely
energetic and chaotic environment, fitting for generating a stochastic
background of gravitational waves resulting from the
superposition of numerous random physical processes. This
gravitational-wave background contains direct information about the
origin of the universe, information inaccessible by any other means of
observation. Let us examine a few points related to these waves.

The generation of gravitational waves during the Big Bang is
an extremely important subject, as it helps us understand how the
universe began and evolved. Shortly after the Big Bang, the universe
was in an extremely hot, dense, and dynamic state, and
several
physical processes that occurred at that time can produce
gravitational waves:
 1. Quantum fluctuations of the inflaton field - During cosmic
    inflation, an extremely brief and violent phase of exponential
    expansion, it is believed that quantum fluctuations in the
    inflaton field, the field responsible for inflation, were stretched
    to macroscopic scales. These fluctuations were the seeds of the
    structure of the universe, which later became galaxies and galaxy
    clusters, and they additionally generated primordial gravitational
    waves. These waves, produced when the universe was between
    $10^{-36}$
and 
$10^{-32}$ seconds old, possess a continuous and
    broad spectrum. As the universe expanded, the wavelengths of these
    oscillations were stretched, pushing them into the
    low-frequency range, invisible to current detectors but possibly
    accessible through indirect signals. Indeed, these waves may leave
    imprints on the cosmic microwave background, especially in the
    B-mode polarization, specific distortions in the polarization
    pattern of the light that has reached us from the surface of last
    scattering, 380,000 years after the Big Bang. Experiments such as
    BICEP/Keck and others aim to search for these indirect signatures,
    which would act as fossils of the inflationary epoch and confirm
    the theory of inflation.
    2. Phase transitions in the primordial universe - Just as water can
    change state, from liquid to vapor for example, the universe can
    also undergo phase transitions associated with spontaneous
    symmetry breaking, such as the separation of the electroweak force
    from the other fundamental forces. These transitions may have
    produced bubbles of a new physical state, which, upon colliding
    and coalescing, would generate gravitational waves.
    3. Topological defects - Some theories predict the formation of
    cosmic strings, domain walls, or magnetic monopoles, which are
    structures left over also from likely phase transitions. Oscillations or
    collisions of these structures would also be sources of stochastic
    gravitational waves.

Gravitational waves prior to the Big Bang are also a possibility. Some
speculative cosmological models go further and suggest that
gravitational waves could have originated in a pre-Big Bang era. Among
these proposals, two models stand out:
    1. Brane theories - In some contexts,
    in particular in the context of string theory, our universe
    could be a brane, a three-dimensional membrane, floating in a
    higher-dimensional space. Collisions between branes, or
    oscillations of these structures within extra dimensions, could
    produce detectable gravitational waves.
    2. Cyclic or bouncing universes - Models such as the ekpyrotic
    scenario or bounce cosmologies propose that the universe underwent
    a contracting phase before the Big Bang, followed by a bounce that
    initiated the present expansion. During the contraction or the
    bounce,  gravitational waves could have been generated with
    properties distinct from those predicted by inflation.

Thus, primordial gravitational waves, both from the Big Bang and from
possible earlier eras, are among the most fascinating frontiers of
modern cosmology. They provide information about the very first
instants of the universe, possibly reaching back to the moment of the
Big Bang itself. Their detection, whether direct or indirect, will
open an entirely new window on physics. As gravitational-wave
observatories evolve, both terrestrial and space-based, the prospect
of hearing the gravitational echo from when the universe was extremely
young, or even from the beginning of time,
is becoming increasingly
realistic. We eagerly await that moment.

\section{Other projects}

\subsection{Other projects along the same path}

\subsubsection{Terrestrial detectors}

LIGO has undergone constant development over the years with the aim of
increasing its sensitivity and detection capability. It is currently
part of a global network of gravitational-wave observatories that
includes the two LIGO detectors in the United States, the Virgo
detector in Italy, and KAGRA in Japan. This LIGO-Virgo-KAGRA
collaboration is part
of Advanced LIGO, i.e., LIGO A. The collaboration has enabled increasingly
precise detections, from the O1 campaign through O4, which ended
recently. The O5 campaign begins in 2027. A
\hskip -0.03cm
proposed
LIGO
A+,
\hskip -0.03cm
a
\hskip -0.03cm
more
\hskip -0.03cm
advanced
\hskip -0.03cm
LIGO,
\hskip -0.03cm
aims
\hskip -0.03cm
to
\hskip -0.03cm
detect
\hskip -0.03cm
more
\hskip -0.03cm
distant
\hskip -0.03cm
and
\hskip -0.03cm
weaker
\hskip -0.03cm
objects.

The next generation of observatories is already being designed and
promises to revolutionize gravitational astronomy even
further. Following LIGO and using terrestrial Michelson
interferometers, the main developing projects
are now enumerated.

    LIGO Voyager, a next-generation proposal based on upgrading
    existing interferometers to increase sensitivity, especially at
    low frequencies. It is a greatly enhanced version of the current
    detectors.

    Cosmic Explorer, a United States project to build a
    third-generation gravitational-wave detector with far greater
    sensitivity than current instruments. With 40 Km arms, ten times
    longer than LIGO's, Cosmic Explorer aims to observe millions of
    black hole and neutron star mergers, investigate the formation of
    the first black holes and the expansion of the universe, test
    gravity under extreme conditions, explore new fundamental laws of
    physics, and detect weak, long-lasting gravitational waves from
    the most distant and ancient regions of the universe. Construction
    is planned for regions in the United states
    such as Texas or New Mexico,
    with operations expected around 2040.

    Einstein Telescope, a European project to build another
    third-generation detector with sensitivity about ten times greater
    than LIGO, Virgo, and KAGRA. It will be an underground
    interferometer, with three 10 Km arms arranged in a triangular
    configuration, located about 200 meters below ground to reduce
    seismic noise. Scientific goals include observing thousands of
    compact-object mergers per year, detecting continuous and
    primordial gravitational waves, studying the formation of the
    earliest cosmic structures, understanding neutron star physics,
    and testing general relativity. Potential locations include
    Sardinia in Italy or a region near the border between Belgium,
    Germany, and the Netherlands. It is projected to begin operations
    around 2040.

\begin{figure*}[h]
\centering
\includegraphics[scale=0.24]{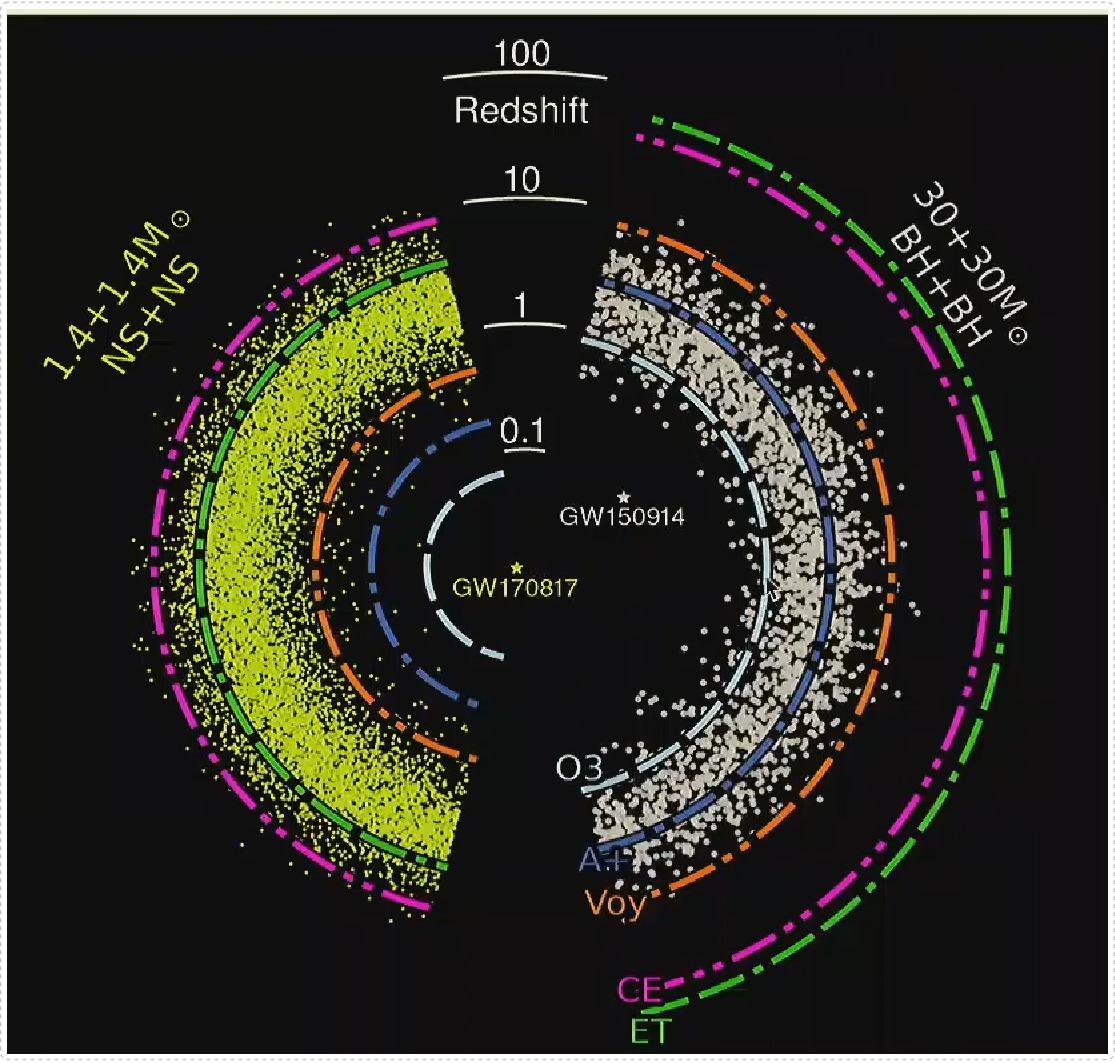}
\caption{An idea of what has already been observed and what will be
observed with ground-based gravitational-wave detectors in the case of
binary systems of compact objects, i.e., neutron stars (NS) and black
holes (BH). The circles indicate the redshift, which is a measure of
distance.
The greater the distance, the further back in time we
observe. The detector
observations shown here are Advanced LIGO during its third
observing run O3, Advanced LIGO with the sensitivity expected for the
fifth observing run, denoted A+, a possible LIGO upgrade called
Voyager, denoted Voy, the Cosmic Explorer, denoted CE, and the Einstein
Telescope, denoted ET. The yellow and white points represent the
statistically predicted population of binary neutron star mergers and
binary black hole mergers, respectively. In terms of distance,
$z=0.1$ corresponds to approximately $400\,{\rm Mpc}$, $z=1$
corresponds to $6000\,{\rm Mpc}$, and $z=10$ and $z=100$ correspond
to distances approaching the edges of the observable universe. The
conversion from parsecs to light-years, for those who prefer this
latter unit, is $1\, {\rm
pc}=3.3\, {\rm light}$-${\rm years}$.
The figure is taken
from a technical report for the Cosmic Explorer project by Evans and
collaborators.}
\label{cosmicexplorerpatrickbrady}
\end{figure*}

In Fig.~\ref{cosmicexplorerpatrickbrady}, 
it is
provided a clear view of what
has already been observed and what is expected to be detected by these
future projects that follow LIGO's line of investigation. These new
observatories promise to deepen our knowledge of the universe by
increasing the reach and sensitivity of gravitational-wave
observations.

\subsubsection{Space-based detectors}

LISA, the Laser Interferometer Space Antenna, is a joint space mission
of ESA with participation from NASA, designed to be the first
space-based gravitational-wave observatory. It will consist of three
spacecraft separated by 2.5 million Km, forming an equilateral
triangle in orbit around the Sun. The satellites act as
interferometers in pairs. Each spacecraft sends and receives laser
beams, measuring tiny changes in distance caused by passing
gravitational waves.
LISA will detect low-frequency gravitational waves, from
$10^{-4}\,{\rm Hz}$ to $1\,{\rm Hz}$, a range inaccessible to
ground-based detectors due to seismic and atmospheric noise. Its
scientific objectives include mergers of supermassive black holes,
compact binary systems of white dwarfs within the Milky Way,
mergers
of intermediate-mass black holes, and signals from the primordial
universe, such as possible echoes of the Big Bang.
It is expected to begin
operating around 2040.
In short, LISA will be essential for understanding the growth of
supermassive black holes, testing general relativity on astronomical
scales, and searching for signals from the primordial universe.

\subsection{Other projects in other directions}

In 2023, a cosmic gravitational-wave background of extremely low
frequency was detected for the first time.
This distant echo is generated by
the superposition of waves produced by mergers of supermassive black
holes throughout time. The
detection became possible by turning the
Milky Way itself into a giant antenna, using a network of millisecond
pulsars distributed across the galaxy. By measuring with extreme
precision the arrival times of signals from these pulsars,
international consortia such as NANOGrav from United States, EPTA
from Europe, InPTA
from India, and PPTA from Australia all part of the general pulsar timing
array PTA project, were able to identify this gravitational
signal. This signal, a stochastic gravitational-wave background, opens
a new avenue for studying galaxy evolution and black holes on
cosmological scales.
With this discovery, new projects are expected to emerge, pushing our
understanding of the universe even further.

\subsection{The gravitational-wave spectrum: sources and detectors}

\begin{figure*}[h]
\centering
\includegraphics[scale=0.3]{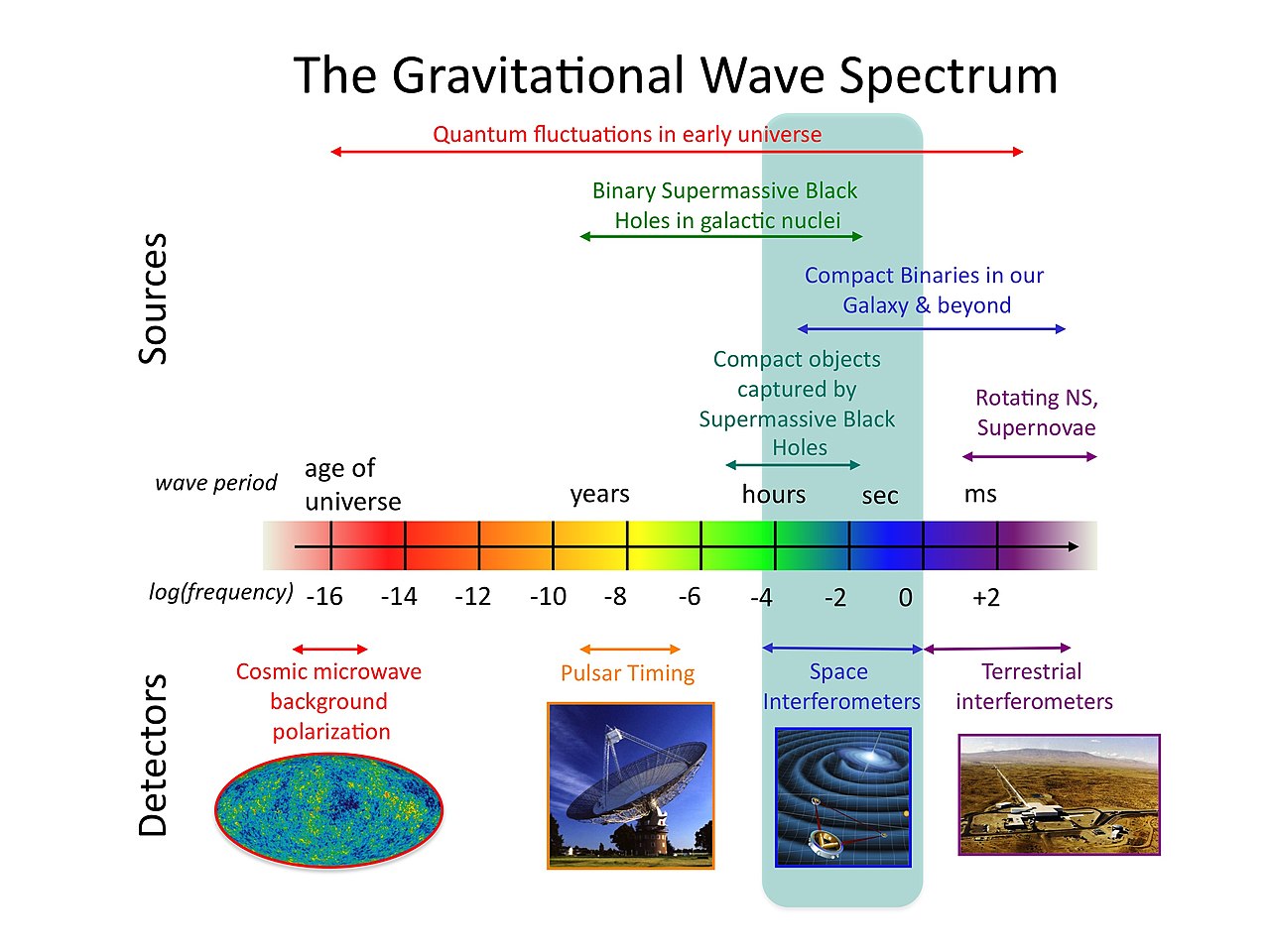}
\caption{The spectrum of gravitational waves with the sources and the
detectors suited to those sources. The figure is taken from an
outreach image by NASA for the LISA mission.}
\label{gwspectrum}
\end{figure*}

The sources of gravitational waves can be many. Indeed, since gravitation
is universal and nothing is completely isolated, anything that moves
emits gravitational waves. But we are interested only in sources that
generate waves detectable with our instruments. For such detection,
the sources must have very large masses, as in neutron stars and black
holes in compact binary systems, and the universe itself. In Fig.
\ref{gwspectrum}
it is shown the main sources of gravitational waves and the detectors
capable of detecting them.

\section{The present and the future}

At present, we are witnessing the birth of a new science:
gravitational-wave astronomy. The first clues are emerging regarding
the formation and interaction of extremely massive objects, such as
stellar-mass black holes and neutron stars. The central mechanisms
behind gamma-ray bursts are gradually being revealed. Furthermore,
progress is being made in investigating the possible contribution of
various types of black holes,
including primordial ones, to solving the
dark-matter problem.

In the not-too-distant future,
perhaps within the next three decades, we
may detect, in real time, mergers of supermassive black holes
resulting from collisions of entire galaxies. It is even possible that
we will capture the echo of gravitational waves generated during the
earliest instants of the Big Bang itself. We will then be facing a
revolution in our observation of the universe.

However, one central question remains open: Will it be possible to
take a decisive step in fundamental physics and fully understand
gravity in the strong-field regime? In other words, is general
relativity valid at all scales except possibly at the Planck scale, or
is the strong-field regime governed by a modified theory of gravity?

\section{Conclusions}

The detection of electromagnetic waves across the entire spectrum
occurred gradually.  Radio waves were detected by Hertz, a German
physicist, in 1886 during experiments carried out to confirm Maxwell's
electromagnetic theory. Infrared
radiation was identified by Herschel, an English astronomer, in 1800,
when he observed the heating caused by sunlight beyond the visible
red. Visible light does not have a specific moment of discovery in
human history: we have always perceived it. Our eyes are, in practice,
natural detectors of this form of electromagnetic radiation, though it
is worth recalling that it was Newton who first realized that white
light is composed of seven colors by passing sunlight through a
prism, an insight he published in 1672, laying the foundations of
modern optics.
Ultraviolet radiation was discovered by Ritter, a German physicist, in
1801 while investigating the effects of sunlight beyond the
violet. X-rays were discovered in 1895 by Röntgen, also a German
physicist, who observed an unknown radiation capable of penetrating
opaque materials. Finally, gamma rays were identified by Villard, a
French chemist, in 1900 while examining the radiation emitted by
radioactive substances.
Gravity waves mixed with matter have been known since the
earliest days of humanity. Ocean waves are, in this sense, a
gravity wave in water. An earthquake can be understood as a
gravity wave propagating through the Earth's crust. Both depend
on the gravitational constant $g$, which governs gravity at the Earth's
surface and is essential for the existence of such
mechanic-gravity waves.
Pure gravitational waves, on the other hand, direct manifestations of
the vibration of spacetime predicted by the theory of general
relativity, depending only on the universal gravitational constant $G$
and the speed of light $c$,
were detected for the first time in human
history on the landmark date of 14 September 2015. On that day, LIGO
recorded the passage of these waves, originating from the merger of
two distant black holes.

\section*{Acknowledgements}

I am grateful to Carlos Herdeiro and Jorge Rocha for their
collaboration and tireless work as editors of the commemorative issue
of Gazeta de Física dedicated to the
10th anniversary of the first
detection of gravitational
waves. I also thank Bernardo Almeida, the journal's
director, for enthusiastically supporting the idea of such a special
issue and for giving me the opportunity to contribute a long-form
article. My thanks go as well to Amaro Rica da Silva for his
invaluable assistance in preparing the manuscript, including help with
LaTeX sources for .doc format  and help in producing figures with
Mathematica to the original
article in Gazeta.

I acknowledge the Fundação para a Ciência e
Tecnologia (FCT) for its financial support through project
UIDB/00099/2025.

\section*{Note}

This article is my English translation of the chapter I wrote for
the special issue of the Portuguese Physics Society journal, Gazeta de
Física, celebrating the 10th anniversary of the first
detection of gravitational
waves. The full reference is J. P. S. Lemos, ``Quando o
espaço-tempo vibra: uma introdução às ondas gravitacionais'', Gazeta
de Física 48(2), 2 (2025).

The article is based on the colloquia I have given since 2017 on the
science of gravitational waves and the methods developed for their
detection. Particularly noteworthy are the colloquium presented at the
Department of Physics of Instituto Superior Técnico,``The Nobel Prize
in Physics 2017: Gravitational Waves and the LIGO Detector'', held on
18 October 2017, immediately after the Nobel Prize announcement, and
the lecture delivered at the XIX Brazilian School of Cosmology and
Gravitation at the Brazilian Center for Physics Research, CBPF, in Rio de
Janeiro, titled ``Black Holes and Gravitational Waves Part II:
Gravitational Waves'', on 17 September 2024.

Below is a bibliography recommended for deepening the understanding of
the topics addressed in this article. The scientific outreach articles
and books included are particularly interesting and offer accessible
reading. The textbooks containing chapters on gravitational waves are
all of excellent quality. Ryder's book was especially useful for this
work, part of our analysis of gravitational waves was adapted from
it. The article announcing the detection of the first gravitational
wave deserves, at the very least, careful consultation, as does the
corresponding official Nobel Prize press release, both essential for
understanding the historical significance of the discovery. This
bibliography is not exhaustive, but it provides a foundation for those
wishing to explore the topic in greater depth.

\section*{References}\label{bibwaves}

\noindent
{\bf
Popular-science articles on gravitational waves:}
\vskip 0.1cm

\noindent
[1] J. D. Toniato, ``What are gravitational waves?'', Cadernos de
Astronomia 2(2), 6 (2021).

\noindent
[2] S. Speziale and D. A. Steer, ``An introduction to gravitational wave
theory'', in 4th MaNiTou Summer School on Gravitational Waves (CNRS
Contemporary Encyclopaedia Sciences, ISTE, London, 2025);
arXiv:2508.21817 [astro-ph].

\noindent
[3] C. A. R. Herdeiro, J. P. S. Lemos, and J. V. Rocha
(editors), {\it 10º Aniversário da 1ª Deteção
de Ondas Gravitacionais},
Gazeta de Física 48(2), 1-85 (2025), in Portuguese.

\vskip 0.3cm

\noindent
{\bf
Popular-science books on gravitational waves:
}
\vskip 0.1cm

\noindent
[4] M. Bartusiak, {\it Einstein's Unfinished Symphony: The Story of a
Gamble, Two Black Holes, and a New Age of Astronomy}
(Yale University
Press, New Haven, 2017).

\noindent
[5] H. Collins, {\it Gravity's Kiss: The Detection of Gravitational Waves}
(MIT Press, Cambridge, Massachusetts, 2017).

\noindent
[6] G. Schilling, {\it Ripples in Spacetime: Einstein, Gravitational Waves,
and the Future of Astronomy} (The Belknap Press of Harvard University
Press, Cambridge, Massachusetts, 2017).

\vskip 0.3cm

\noindent
{\bf
Textbooks with chapters on
gravitational waves:}
\vskip 0.1cm

\noindent
[7] C. W. Misner, K. S. Thorne, and J. A. Wheeler, {\it Gravitation} (Freeman,
San Francisco, 1973); new edition (Princeton University Press,
Princeton, 2018).

\noindent
[8] B. F. Schutz, {\it A First Course in General Relativity} (Cambridge
University Press, Cambridge, 1985); new edition (Cambridge University
Press, Cambridge, 2022).

\noindent
[9] M. Maggiore, {\it Gravitational Waves. Volume 1: Theory and Experiment}
(Oxford University Press, Oxford, 2007).

\noindent
[10] L. Ryder, {\it Introduction to General Relativity} (Cambridge
University Press, Cambridge, 2009).

\vskip 0.3cm

\noindent
{\bf
The paper that announced the
first detection of gravitational waves:}
\vskip 0.1cm

\noindent
[11] B. P. Abbott et al. (LIGO Scientific Collaboration and Virgo
Collaboration), ``Observation of gravitational waves from a binary
black hole merger'', Physical Review Letters 116, 061102 (2016);
arXiv:1602.03837 [gr-qc].

\vskip 0.3cm

\noindent
{\bf
The 2017 Nobel Prize in Physics for
gravitational waves:}
\vskip 0.1cm

\noindent
[12] The Nobel Committee for Physics, ``Scientific background: The
laser interferometer gravitational-wave observatory and the first
direct observation of gravitational waves'', Advanced Information,
Royal Swedish Academy of Sciences (2017);
https://www.nobelprize.org/prizes/physics/2017/advanced-information/.

\noindent
[13]
J. P. S. Lemos, ``The Nobel Prizes in Physics for astrophysics and
gravitation and the Nobel Prize for black holes: Past, present, and
future'', Gazeta de Física 44(2/3), 58 (2021); arXiv:2112.14346
[physics.hist-ph] (2021). The paper in Gazeta de Física is
in Portuguese with title ``Os prémios Nobel de física para
astrofísica e gravitação e o prémio Nobel
para buracos negros: Passado, presente
e futuro''. The paper in the arXivs is its translation to English.

\end{document}